\documentclass[aps,a4paper,twocolumn]{revtex4}
\usepackage{amsmath}
\usepackage{graphicx}
\usepackage{subfigure}
\usepackage[usenames,dvipsnames]{color}
\definecolor{darkblue}{RGB}{0,0,196}
\usepackage[colorlinks=true,linkcolor=darkblue,citecolor=darkblue,urlcolor=darkblue]{hyperref}
\usepackage{setspace}
\usepackage{hyperref}
\usepackage{xcolor}
\hypersetup{
  colorlinks   = true, 
  urlcolor     = red, 
  linkcolor    = blue, 
  citecolor   = blue 
}
\usepackage{graphicx}
\usepackage{amsmath,bbm}
\usepackage{amssymb,bm}
\usepackage{yfonts}
\usepackage{comment}

\begin{document}
\title{Probing initial geometrical anisotropy and final azimuthal anisotropy in heavy-ion collisions at Large Hadron Collider energies through event-shape
engineering}
\author{Suraj Prasad$^{1}$}
\author{Neelkamal Mallick$^{1}$}
\author{Sushanta Tripathy$^{2}$}
\author{Raghunath Sahoo$^{1}$\footnote{Corresponding Author email: Raghunath.Sahoo@cern.ch}}
\affiliation{$^{1}$Department of Physics, Indian Institute of Technology Indore, Simrol, Indore 453552, India}
\affiliation{$^{2}$INFN - sezione di Bologna, via Irnerio 46, 40126 Bologna BO, Italy}

\begin{abstract}
\noindent

Anisotropic flow is accredited to have effects from the initial state geometry and fluctuations in the nuclear overlap region. The elliptic flow ($v_2$) and triangular flow ($v_3$) coefficients of the final state particles are expected to have influenced by eccentricity ($\varepsilon_2$) and triangularity ($\varepsilon_3$) of the participants, respectively. In this work, we study $v_2$, $v_3$, $\varepsilon_2$, $\varepsilon_3$ and the correlations among them with respect to event topology in the framework of a multi-phase transport model (AMPT). We use transverse spherocity and reduced flow vector as event shape classifiers in this study. Transverse spherocity has the unique ability to separate events based on geometrical shapes, i.e., jetty and isotropic, which pertain to
pQCD and non-pQCD domains of particle production in high-energy physics, respectively. We use the two-particle correlation method to study different anisotropic flow coefficients. We confront transverse spherocity with a more widely used event shape classifier -- reduced flow vector  ($q_n$) and they are found to have significant (anti-)correlations among them. We observe significant spherocity dependence on $v_2$, $v_3$ and $\varepsilon_2$. This work also addresses transverse momentum dependent crossing points between $v_2$ and $v_3$, which varies for different centrality and spherocity percentiles.

\end{abstract}
\date{\today}
\maketitle  

\section{Introduction}
\label{intro}
The goal of ultra-relativistic heavy-ion collisions is to produce a thermalised deconfined medium of quarks and gluons that existed shortly after the Big Bang. These deconfined partons in thermal equilibrium is well known as quark-gluon plasma (QGP). Since QGP can not be detected directly, several indirect signatures have been proposed to signify the formation of a hot and dense medium. Proton-proton (pp) collisions are traditionally used as a baseline to study some of these signatures in heavy-ion collisions. However, recently, some of the heavy-ion-like signatures of QGP have been observed in pp collisions \cite{ALICE:2016fzo, Khuntia:2018znt, CMS:2016fnw, Bjorken:2013boa} which compelled the scientific community to investigate pp collisions from different aspects using state of the art methods. Transverse spherocity is one such observable which separates events based on geometrical shapes, such as jetty and isotropic events. So far, transverse spherocity is broadly explored in pp collisions \cite{Cuautle:2014yda,Cuautle:2015kra,Salam:2009jx,Bencedi:2018ctm,Banfi:2010xy,Tripathy:2019blo,Khuntia:2018qox,Deb:2020ezw,Khatun:2019dml}; however, it is still novel for heavy-ion collisions. A recent study \cite{Prasad:2021bdq} of transverse spherocity in heavy-ion collisions shows that many of the global observables in heavy-ion collisions such as kinetic freeze-out temperature, mean transverse radial flow velocity, mean transverse mass, integrated yield, etc., strongly correlate with transverse spherocity.

\begin{figure}
\includegraphics[scale=0.42]{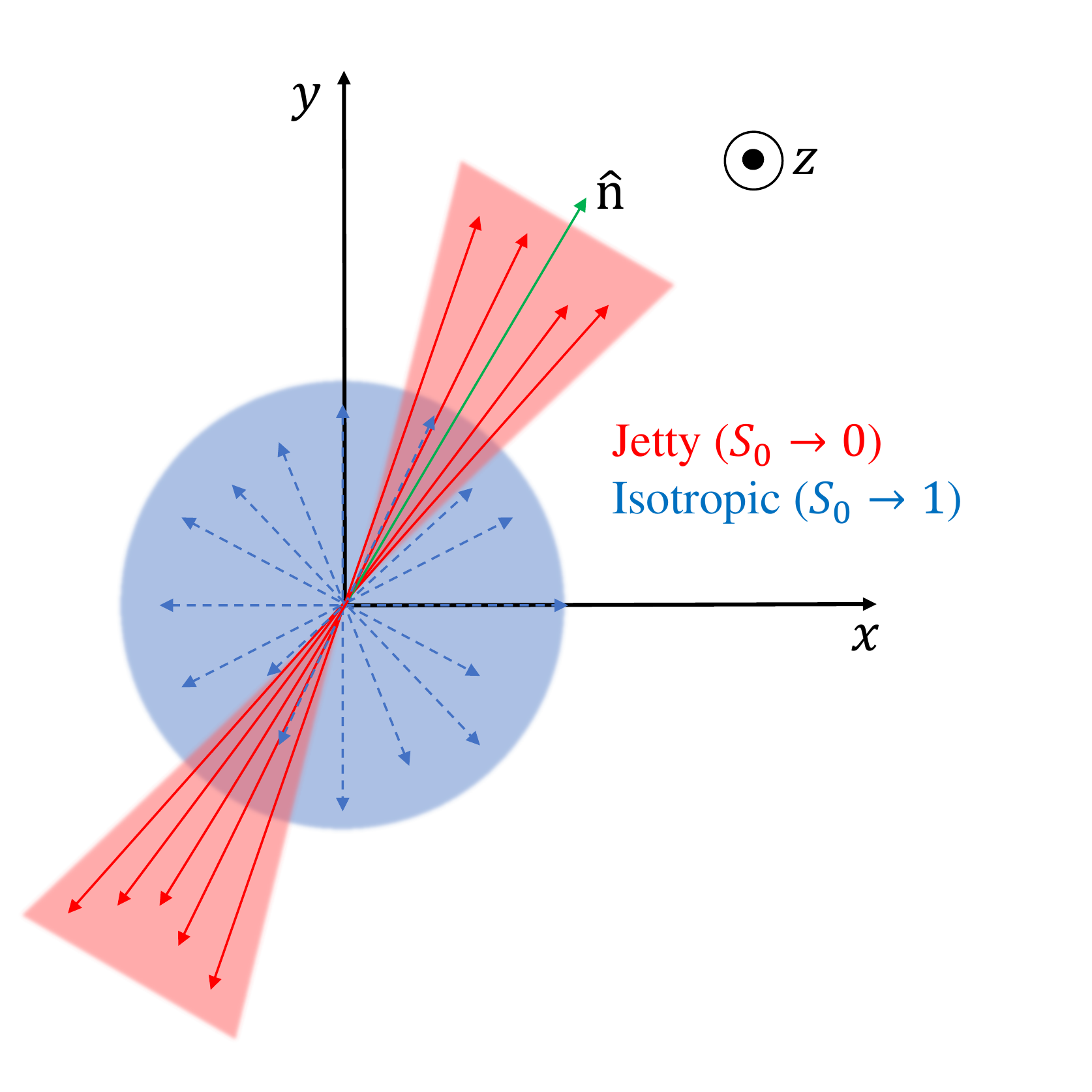}
\caption[width=18cm]{(Color Online) Schematic picture showing jetty and isotropic events in the transverse plane.}
\label{fig:spherosch}
\end{figure}

The anisotropic flow is related to momentum space azimuthal anisotropy and it is parameterised by the coefficients of Fourier expansion of momentum distribution. These anisotropic flow coefficients are closely associated with initial geometry as well as its fluctuations and the equation of state of the medium formed. QGP behaves like a perfect fluid, and anisotropic flow studies at Relativistic Heavy Ion Collider (RHIC) and Large Hadron Collider (LHC) energies show that it has a viscosity to entropy density ratio ($\eta/s$) close to $1/4\pi$ \cite{Schenke:2011bn, Gale:2012rq}, which is the lower bound as imposed by the quantum mechanical considerations based on supersymmetric gauge theory in infinite coupling limit \cite{Danielewicz:1984ww,Kovtun:2004de}. 
With an increase in the order of flow coefficients, their sensitivity to $\eta/s$ increases, i.e., triangular flow is more sensitive to $\eta/s$ than elliptic flow \cite{Schenke:2011bn}. Using flow coefficients, one can infer the fluidity of the medium formed; for example, a lower value of the second-order flow coefficient, elliptic flow, corresponds to a relatively higher value of $\eta/s$ and vice versa \cite{Schenke:2011bn}. The comparison of hydrodynamic studies with the experimentally measured anisotropic flow values shows that the major contribution of the anisotropic flow is expected to arise from the partonic medium and evolve with the evolution of QGP \cite{ALICE:2014wao,BRAHMS:2004adc,Shen:2015msa,PHOBOS:2004zne,STAR:2005gfr}. However, this is not all to the anisotropic flow, because it also has effects from the hadronic rescattering \cite{Hirano:2005xf,Pradhan:2021zbt}. These contributions lead to the number of constituent quarks (NCQ) scaling hierarchy of different particles in anisotropic flow at intermediate transverse momentum ($p_{\rm T}$) range, i.e., baryons have more flow compared to mesons \cite{Voloshin:2002wa}. However, this NCQ scaling on elliptic flow is violated at LHC energies \cite{ALICE:2014wao, ALICE:2010suc}. In recent studies~\cite{Mallick:2020ium, Mallick:2021hcs}, it has been observed that elliptic flow anti-correlates with transverse spherocity \cite{Mallick:2020ium} i.e., events with higher values of spherocity (isotropic events) have lower elliptic flow and vice versa. This result is expected, since high spherocity events are isotropic in nature, thus its momentum space azimuthal anisotropy is expected to be less. A study on the NCQ scaling of elliptic flow at LHC energies shows that the scaling is violated for the integrated and jetty types of events \cite{Mallick:2021hcs}. So far the sensitivity of transverse spherocity on higher harmonic flow coefficients and initial state geometrical anisotropy are yet to be investigated. 

This study aims to address the dependence of transverse spherocity on eccentricity, triangularity, elliptic and triangular flow using a multi-phase transport model (AMPT). We also perform a detailed study on the correlation of transverse spherocity with more traditional event shape observable, i.e, reduced flow vector~ \cite{STAR:2002hbo}.

This paper is organised as follows. We begin with a brief introduction and motivation of the study in section \ref{intro}. Then we discuss the event generation and analysis methodology in section \ref{section2}, where we introduce AMPT and transverse spherocity. In section \ref{section3}, we define the formulations of eccentricity, triangularity, elliptic flow, and triangular flow and discuss our results. Finally, in section \ref{section4}, we summarise our results with important findings.

\section{Event Generation and Analysis Methodology}
\label{section2}
In this section, we briefly discuss the event generator i.e., a multi-phase transport model which is used in this study. Then we discuss and compare the event classifiers used for event topology analysis.

\subsection{A Multi-Phase Transport (AMPT) Model}
\label{section:ampt}
AMPT model includes both initial partonic collisions and final hadronic interactions and the transition between these two phases of matter. It has four main components, namely, initialization of collisions, parton transport after initialization, hadronization mechanism and hadron transport  \cite{Zhang:1999mqa,Zhang:2000bd,Zhang:2000nc,Lin:2001cx,Lin:2001yd,Pal:2001zw,Lin:2001zk,Zhang:2002ug,Pal:2002aw,Lin:2002gc,Lin:2003ah,Lin:2003iq,Wang:1991hta,Zhang:1997ej,He:2017tla,Li:2001xh,Greco:2003mm}. A brief discussion on the main components of AMPT can be found in the appendix section. Since, the particle flow and spectra are well described at mid-$p_{\rm T}$ region by quark coalescence mechanism for hadronisation \cite{Fries:2003vb,Fries:2003kq}, we have used AMPT string melting (SM) mode (AMPT version 2.26t9b) for our study. The AMPT settings in the current study are the same as reported in Ref. \cite{Tripathy:2018bib}. Centrality selection has been done using impact parameter slicing and we have used the impact parameter values for different centralities from Ref. \cite{Loizides:2017ack} for our current analysis.

\subsection{Event shape observables}
\label{section:eventshape}

\subsubsection{Transverse spherocity ($S_0$)}
\label{section:spherocity}

Transverse spherocity ($S_0$) is an event topology classifier that is used to separate events based on geometrical shapes. Transverse spherocity can be defined for a unit vector $\hat{n} (n_T, 0)$ as follows:
\begin{equation}
    S_0=\frac{\pi^2}{4} \text{min}\Big(\frac{\sum_{i} |\vec{p_{T_i}}\times\hat{n}|}{\sum_{i}|\vec{p_{T_i}}|}\Big)^2
\label{eqn:sphero}    
\end{equation}
where $p_{T_i}$ denotes transverse momentum of hadron $i$ and $i$ runs over all the final state hadrons in an event. $\hat{n}$ is chosen such that the bracketed term in Eq. \ref{eqn:sphero} is minimised. This selection is done by iterating through all the possible values of $\hat{n}$ in the transverse plane for the event. Multiplication of $\pi^2/4$ in Eq. \ref{eqn:sphero} ensures that $S_0$ is normalised and ranges between 0 and 1. The two extreme limits of $S_0$, namely, 0 and 1 correspond to the jetty (pencil-like) and isotropic events, respectively. Figure~\ref{fig:spherosch} represents a schematic picture showing jetty and isotropic events in the transverse plane. In order to create similar conditions as in ALICE experiment at the LHC, we only select particles with $|\eta|<0.8$ and $p_{\rm{T}}>$~0.15~GeV/$c$ with a minimum constraint of 5 charged particles in a collision. For the sake of simplicity, in this paper, we may sometimes refer transverse spherocity as spherocity. Here we separate events based on spherocity i.e., by choosing extreme $20\%$ events from the spherocity distribution as done in Refs.\cite{Prasad:2021bdq, Mallick:2020ium}. We call the events having highest and lowest $20\%$ values in the $S_0$ values as high-$S_0$ and low-$S_0$ events, respectively. The spherocity cuts for high-$S_0$ and low-$S_0$ events are given in Table \ref{tab:sphero}. In a recent study, $v_2$ is found to be strongly anticorrelated with $S_0$ \cite{Mallick:2020ium}. This motivates us to look for the correlations among the spherocity and more traditional event shape classifier in heavy-ion collisions, i.e., the reduced flow vector ($q_n$). 

\begin{table}[ht!]
\caption{Low-$S_0$ and high-$S_0$ cuts on spherocity distribution for different centrality classes in Pb--Pb collisions at $\sqrt{s_{\rm{NN}}} = 5.02$~TeV \cite{Prasad:2021bdq, Mallick:2020ium}. }
\label{tab:sphero}
\begin{tabular}{ |c|p{2cm}|p{2cm}|}
\hline
Centrality (\%) & Low-$S_{0}$ & High-$S_{0}$\\
\hline
0-10        & 0 -- 0.880      & 0.953 -- 1 \\
10-20       & 0 -- 0.813      & 0.914 -- 1 \\
20-30		& 0 -- 0.760	  &	0.882 -- 1 \\
30-40		& 0 -- 0.735	  &	0.869 -- 1 \\
40-50		& 0 -- 0.716	  &	0.865 -- 1 \\
50-60		& 0 -- 0.710	  &	0.870 -- 1 \\
60-70		& 0 -- 0.707	  &	0.873 -- 1 \\	
\hline
\end{tabular}
\end{table}

\begin{figure*}
\includegraphics[scale=0.8]{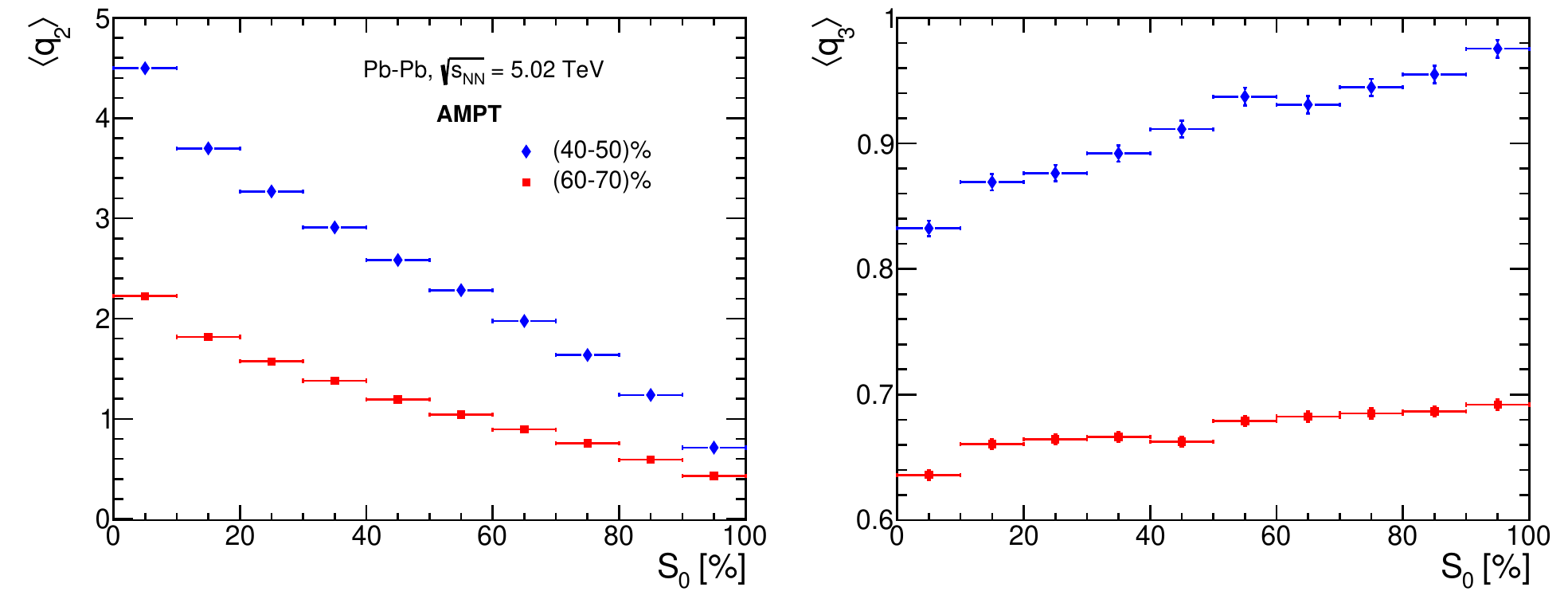}
\caption[width=18cm]{(Color Online) Event averaged $q_2$ (left) and $q_3$ (right) vs transverse spherocity ($S_0$) for  midcentral (40-50)\%, and peripheral (60-70)\% Pb--Pb collisions at $\rm\sqrt{s_{NN}}$ = 5.02 TeV using AMPT model.}
\label{fig:qnsphero}
\end{figure*}

\subsubsection{Reduced flow vector ($q_n$)}
\label{section:qn}

Another event classifier, called as reduced flow vector ($q_{\rm n}$) is traditionally used in heavy-ion collisions to perform event shape engineering. The magnitude of reduced flow vector of order-$n$ \cite{STAR:2002hbo} is given as,
\begin{equation}
    q_n= \frac{|Q_n|}{\sqrt{M}}
\end{equation}
where,
\begin{equation}
    Q_n= \sum_{j=1}^{M}e^{in\phi_{j}} ,
\end{equation}
$\phi_j$ is the azimuthal angle of $j^{\rm th}$ particle at the kinetic freeze-out, $M$ is the multiplicity of the event, and $i$ is the imaginary unit number.

In the limit, $M\to\infty$, $q_2$ approaches transverse energy ($E_T$) weighted single particle elliptic flow ($v_2$) \cite{Schukraft:2012ah, ATLAS:2015qwl}. $q_n$ are found to be strongly correlated with $v_n$ values as shown in Ref. \cite{ATLAS:2015qwl}. For the sake of comparison of $q_n$ with transverse spherocity in the same footing, similar $\eta$ and $p_{\rm T}$ cuts have been applied in $q_n$ calculations as done in transverse spherocity. The left panel of Fig.~\ref{fig:qnsphero} represents the correlation between spherocity and $q_2$, where both of them are observed to be anti-correlated, i.e., high-$S_0$ events have lower $q_2$ values and vice versa. However, in the right panel of Fig.~\ref{fig:qnsphero}, where $\langle q_3\rangle$ is shown against different $S_0$ selections, we observe a mild positive correlation of $\langle q_3\rangle$ with $S_0$. This correlation between spherocity and $q_3$, indicates a finite dependence of $v_3$ on spherocity \cite{ATLAS:2015qwl}.

\section{Results and Discussions}
\label{section3}
In this section, we present a detailed discussion on the spherocity dependence of the geometry of the nuclear overlap region, i.e., eccentricity and triangularity ( \ref{section3.1}). Then we discuss the methodology and results on coefficients of anisotropic flow and the interplay among them as a function of spherocity, presented in section~\ref{section3.2}.

\subsection{Eccentricity and Triangularity}
\label{section3.1}

The overlap region in a non-central heavy-ion collision is not isotropic in space. If the pressure gradient of the hot and dense medium formed in the heavy-ion collisions is large enough, the anisotropy in initial geometry can be translated into final momentum space azimuthal anisotropy. The anisotropy in the initial spatial distribution of the nucleons in the overlap region can be quantified by the quantities such as eccentricity ($\varepsilon_2$), triangularity ($\varepsilon_3$) etc. As the name suggests, eccentricity refers to how elliptical the medium can be, and is generated due to anisotropy in the nuclear overlap region. Similarly, triangularity characterises the triangular geometry of the overlap region during the collision of two heavy-ions and arises due to event-by-event fluctuations in the participant nucleon collision points \cite{Alver:2010gr}.
In the current study, we have used AMPT model to see the dependence of transverse spherocity on eccentricity ($\rm{\varepsilon_{2}}$) and triangularity ($\rm{\varepsilon_{3}}$) having followed the notations used in Ref. \cite{Petersen:2010cw}.
Eccentricity and triangularity of the participant nuclei can be generalised as follows \cite{Petersen:2010cw}:
\begin{equation}
\rm\varepsilon_{n}=\frac{\sqrt{\langle{r^{n}cos(n\phi_{\text{part}})}\rangle^{2}+\langle{r^{n}\sin(n\phi_{\text{part}})}\rangle^{2}}}{\langle{r^{n}}\rangle}
\label{eq:eccentricity}
\end{equation}

where $r$ and $\rm{\phi_{\text{part}}}$ are the polar co-ordinates of the participant nucleons. $\rm n$ = 2 corresponds to eccentricity and $\rm n$ = 3 corresponds to triangularity. A higher value of $\rm n$ can be used to study higher-order spatial anisotropy of the collision overlap region, although the contributions from the higher order terms will be smaller. In Eq. \ref{eq:eccentricity}, angular brackets, ``$\langle ... \rangle$" represents the mean taken over all the participant nucleons in an event.

\begin{figure}
\includegraphics[scale=0.5]{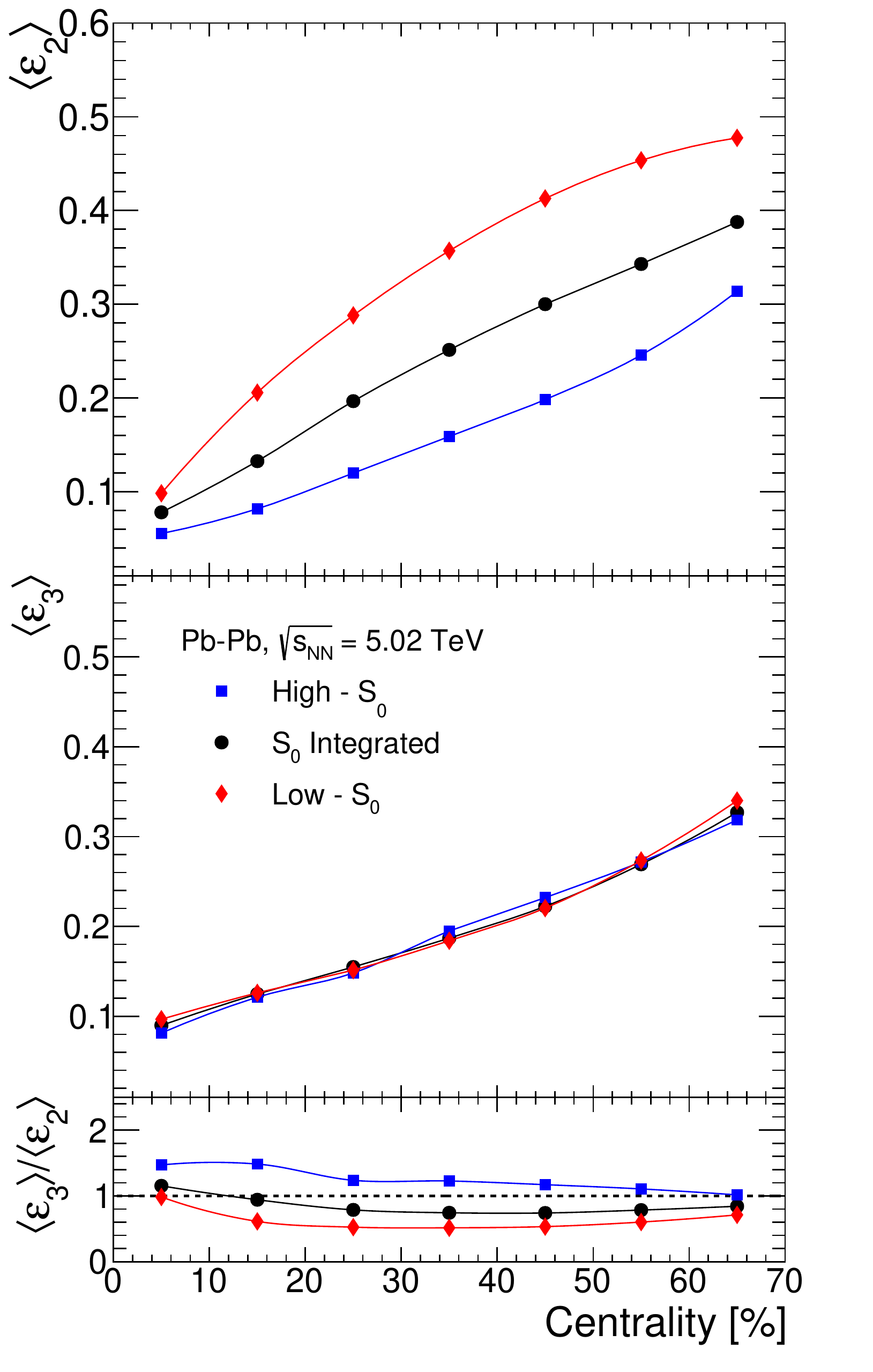}
\caption[width=18cm]{(Color Online)Event-average eccentricity ($\langle\varepsilon_{2}\rangle$) (top), triangularity ($\langle\varepsilon_{3}\rangle$) (middle) and the ratio ($\rm{\langle \varepsilon_{3}\rangle}/\rm{\langle\varepsilon_{2}\rangle}$) (bottom) as a function of centrality for low-$\rm{S_{0}}$ (red diamond), high-$\rm{S_{0}}$ (blue square) and integrated $\rm{S_{0}}$ (black circle) events in Pb--Pb collisions at $\rm{\sqrt{s_{NN}} =  5.02}$  TeV using AMPT. }
\label{fig:ratioe2e3}
\end{figure}

Figure \ref{fig:ratioe2e3} shows the event-average eccentricity ($\langle\varepsilon_{2}\rangle$) (top) and triangularity ($\langle\varepsilon_{3}\rangle$) (middle) and their ratio (bottom) as a function of centrality for different spherocity events. Both $\langle\varepsilon_{2}\rangle$ and $\langle\varepsilon_{3}\rangle$ have clear dependence on the centrality and are increasing from central to peripheral collisions. This behavior is expected since in central collisions, the nuclear overlap region is more spatially symmetric compared to that in peripheral collisions, resulting in low values of $\langle\varepsilon_{2}\rangle$ and $\langle\varepsilon_{3}\rangle$ in central collisions. However, as one moves towards peripheral collisions, the participating nucleon overlap region gets more and more spatially anisotropic resulting in higher values of $\langle\varepsilon_{2}\rangle$ and $\langle\varepsilon_{3}\rangle$.
In a particular centrality, one notices lower $\langle\varepsilon_{2}\rangle$ for high-$\rm S_{0}$ events than the low-$\rm S_{0}$ events. This indicates that transverse spherocity can also be used to distinguish events based on the initial geometry.
However, since the contribution in $\varepsilon_n$ for $n > 2$ arises due to fluctuations in the density profile of the participating nucleons, one should not expect significant transverse spherocity dependence on $v_3$, as shown in the middle plot of Fig.~\ref{fig:ratioe2e3}.
The bottom plot of Fig.~\ref{fig:ratioe2e3} represents the ratio of ($\langle\varepsilon_{3}\rangle$) to ($\langle\varepsilon_{2}\rangle$), plotted against different centrality classes for high-$\rm{S_{0}}$, $\rm{S_{0}}$ integrated and low-$\rm{S_{0}}$ classes. From the ratio, it is clear that $\langle\varepsilon_{2}\rangle$ is higher than $\langle\varepsilon_{3}\rangle$ for low-$\rm{S_{0}}$, and $\rm{S_{0}}$ integrated  events throughout all centrality classes except the most central case where, $\langle\varepsilon_{3}\rangle$ seems to be higher. But the order is reversed for high-$\rm{S_{0}}$ events where $\langle\varepsilon_{3}\rangle$ is always greater than $\langle\varepsilon_{2}\rangle$, indicating the dominance of density fluctuations over the geometry of the overlap region.

\begin{figure}
\includegraphics[scale=0.4]{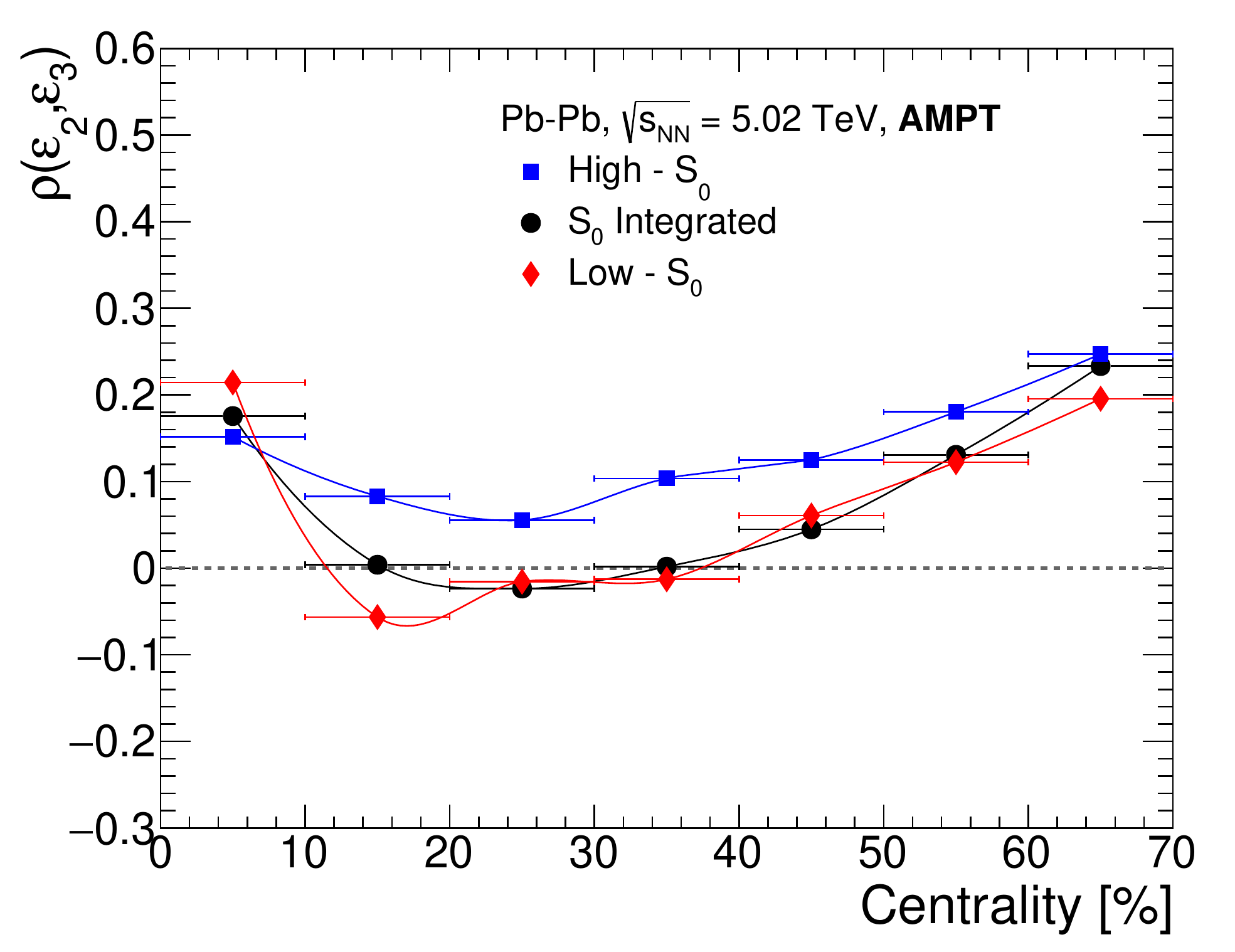}
\caption[width=18cm]{(Color Online) Pearson correlation coefficient between eccentricity and triangularity as a function of centrality for low-$\rm{S_{0}}$ (red diamond), high-$\rm{S_{0}}$ (blue square) and integrated $\rm{S_{0}}$ (black circle) events in Pb--Pb collisions at $\rm{\sqrt{s_{NN}} =  5.02}$  TeV using AMPT. }
\label{fig:pearsoncorrelatione2e3}
\end{figure}

To quantify the correlation between eccentricity and triangularity, we calculate the Pearson correlation coefficient ($\rho(\varepsilon_2, \varepsilon_3)$) using Eq.~\ref{eq:PCC}.

\begin{widetext}
\begin{equation}
\rho(\varepsilon_2, \varepsilon_3)=\frac{N_{\rm ev} \sum_{i=1}^{N_{\rm ev}}\varepsilon_{2,i}\varepsilon_{3,i}-\big(\sum_{i=1}^{N_{\rm ev}}\varepsilon_{2,i}\big)\big(\sum_{i=1}^{N_{\rm ev}}\varepsilon_{3,i}\big)}{\sqrt{\big[N_{ev}\sum_{i=1}^{N_{\rm ev}}\varepsilon_{2,i}^{2}-(\sum_{i=1}^{N_{\rm ev}}\varepsilon_{2,i})^{2}\big]\big[N_{ev}\sum_{i=1}^{N_{\rm ev}}\varepsilon_{3,i}^{2}-(\sum_{i=1}^{N_{\rm ev}}\varepsilon_{3,i})^{2}\big]}}
\label{eq:PCC}
\end{equation}
\end{widetext}
 Where, $N_{ev}$ represents the number of events for a given centrality and spherocity class. $\varepsilon_{2,i}$ and $\varepsilon_{3,i}$ represent the eccentricity and triangularity for the $i$th event, respectively. The value of $\rho(\varepsilon_2, \varepsilon_3)$ lies between -1 to 1; positive value indicates positive correlation while negative value of $\rho(\varepsilon_2, \varepsilon_3)$ implies anti-correlation. Figure~\ref{fig:pearsoncorrelatione2e3} shows the Pearson correlation coefficient between eccentricity and triangularity, denoted as $\rho(\varepsilon_2, \varepsilon_3)$, for different centrality and spherocity classes in Pb--Pb collisions at $\rm{\sqrt{s_{NN}} = 5.02}$ TeV. The correlation appears to be explicitly higher for the most central case, then suddenly decreases for the mid-central cases and starts to rise again towards the peripheral collisions. This peculiar behaviour of the correlation between eccentricity and the triangularity is extended for different spherocity classes, and it is found that for high-$S_{0}$ class of events, the correlation is comparatively higher and positive compared to low-$S_{0}$, and $S_{0}$-integrated events. Even though the transverse spherocity is a final state event shape classifier, it successfully separates the observables related to initial geometry and establishes a correlation between eccentricity and triangularity for the isotropic case. It should also be noted that this correlation could be affected by the initial partonic scatterings, which one finds in the AMPT-SM model \cite{Ma:2016hkg}.

\subsection{Elliptic and Triangular Flow}
\label{section3.2}

Anisotropic flow is a measure of the azimuthal momentum anisotropy of the final state particles produced in a collision. Anisotropic flow depends upon initial spatial anisotropy in the nuclear overlap region, transport properties, and the equation of state of the system. Anisotropic flow can be characterised by the coefficients of the Fourier expansion of momentum distribution of the final state particles and is given by:
\begin{equation}
    E\frac{d^3N}{dp^3}=\frac{d^2N}{2\pi p_{T}dp_{T}dy}\Big(1+2\sum^{\infty}_{n=1}v_{n}\cos[n(\phi-\psi_{n})]\Big)
    \label{eqn-invfourierexp}
\end{equation}
Here $\phi$ is the azimuthal angle of the particles in the transverse plane and $\psi_{n}$ is the nth harmonic event plane angle \cite{ALICE:2014wao}. $n$ stands for the order of the anisotropic flow coefficient. $n$=2 stands for elliptic flow ($v_2$) and $n$=3 refers to the triangular flow ($v_3$).
In general $n$th order anisotropic flow, $v_{n}$ can be defined as:
\begin{equation}
    v_{n}=\langle\cos[n(\phi-\psi_n)]\rangle
    \label{eqn-vnavg}
\end{equation}

We are dealing with different kinds of spherocity events and among them, low-$\rm S_{0}$ events are prone to have contributions from jets. Thus, in order to see the fair dependence of transverse spherocity, we use the two-particle correlation method to study the elliptic and triangular flow as done in Refs. \cite{PHENIX:2008osq,ATLAS:2012at,ATLAS:2015qwl}. The two-particle correlation method has an advantage because it deals with the non-flow effects caused by jets and resonance decays by implementing a proper pseudorapidity cut. It has an additional advantage since it does not require event plane angle ($\psi_n$) to calculate $v_n$.  In experiments, a pseudorapidity dependence of $\psi_n$ is observed, which is not taken into consideration in the present study for simplicity, and it does not affect the performed studies. Here we assume $\psi_n$ to be the global phase angle for all the particles irrespective of the 
selection of the bins in pseudorapidity \cite{Mallick:2020ium,CMS:2015xmx, ATLAS:2017rij}. Although in AMPT we can set 
the reaction plane angle, $\psi_R = 0$, this is not trivial in experiments to determine the same. Thus, one follows a two-particle correlation approach. The steps to find the correlation function is as follows: 
\begin{itemize}
    \item We compose two sets of particles based on particle transverse momentum denoted by labels `a' and `b'.
    \item Each particle from `a' pairs with every particle in `b' and the relative pseudorapidities ($\Delta\eta = \eta_a - \eta_b$) and 
    the relative azimuthal angles ($\Delta\phi = \phi_a - \phi_b$) are then determined.
    \item Two particle correlation function ($C(\Delta\eta,\Delta\phi)$) can be constructing by taking the ratio of the same-event pair ($S(\Delta\eta,\Delta\phi)$) distribution to the mixed-event pair distribution ($B(\Delta\eta,\Delta\phi)$). The ratio improves the pair acceptance and ensures no non-uniformity.
\end{itemize}
In the mixed event background, five events are randomly chosen so that it contains no physical correlation \cite{Mallick:2020ium}. To remove the non-flow contributions and to obtain 1D correlation in $\Delta\phi$ distribution, one uses the $\Delta\eta$ cut in $S(\Delta\eta,\Delta\phi)$ and $B(\Delta\eta,\Delta\phi)$ as $2\leq|\Delta\eta|\leq5$ to get $S(\Delta\phi)$ and $B(\Delta\phi)$, respectively. The ratio of $S(\Delta\phi)$ to $B(\Delta\phi)$ is given by $C(\Delta\phi)$, known as 1D correlation in $\Delta\phi$ distribution. 1D correlation in $\Delta\phi$ distribution can also be written as \cite{ATLAS:2012at,ATLAS:2015qwl}:
\begin{figure}
\includegraphics[scale=0.42]{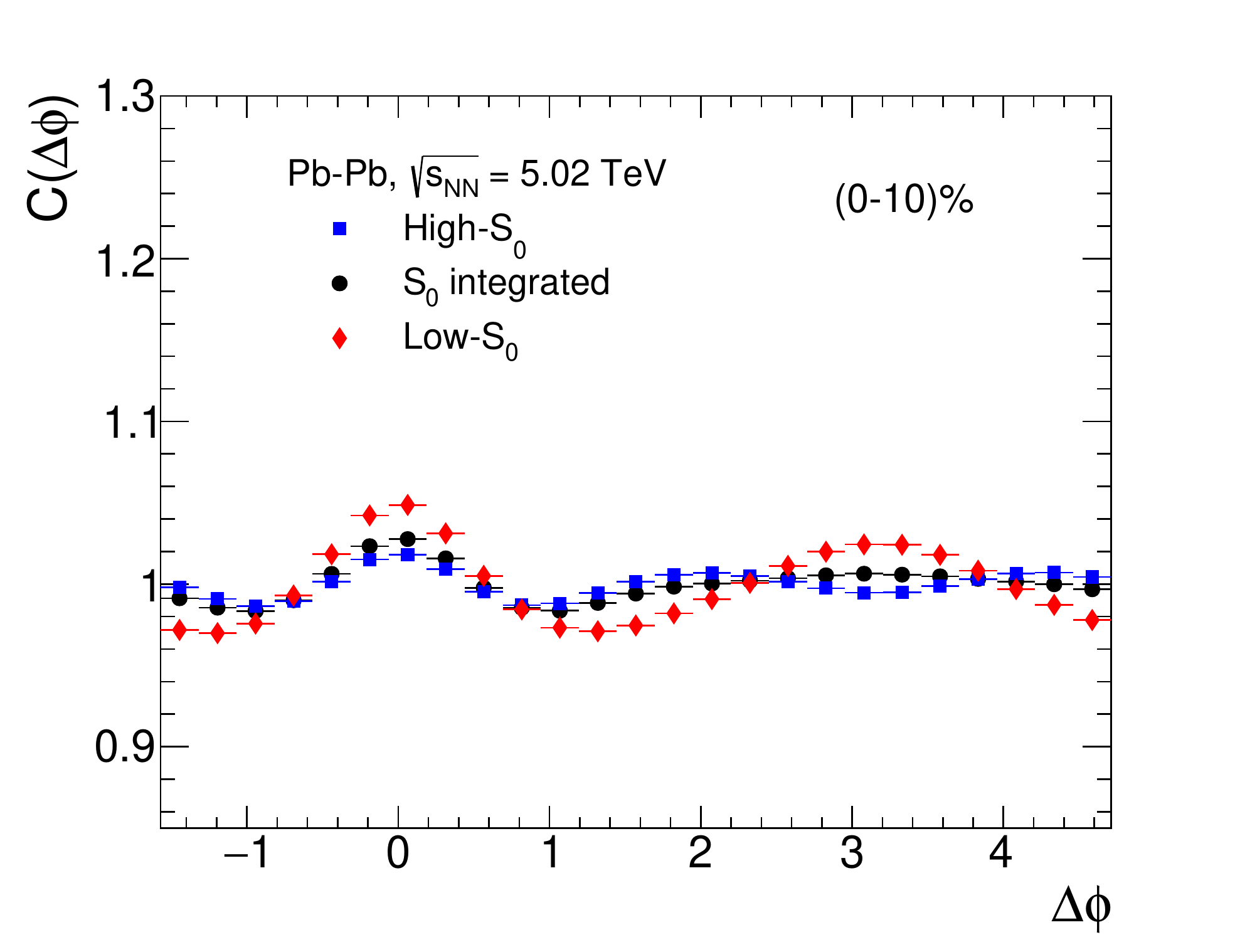}
\includegraphics[scale=0.42]{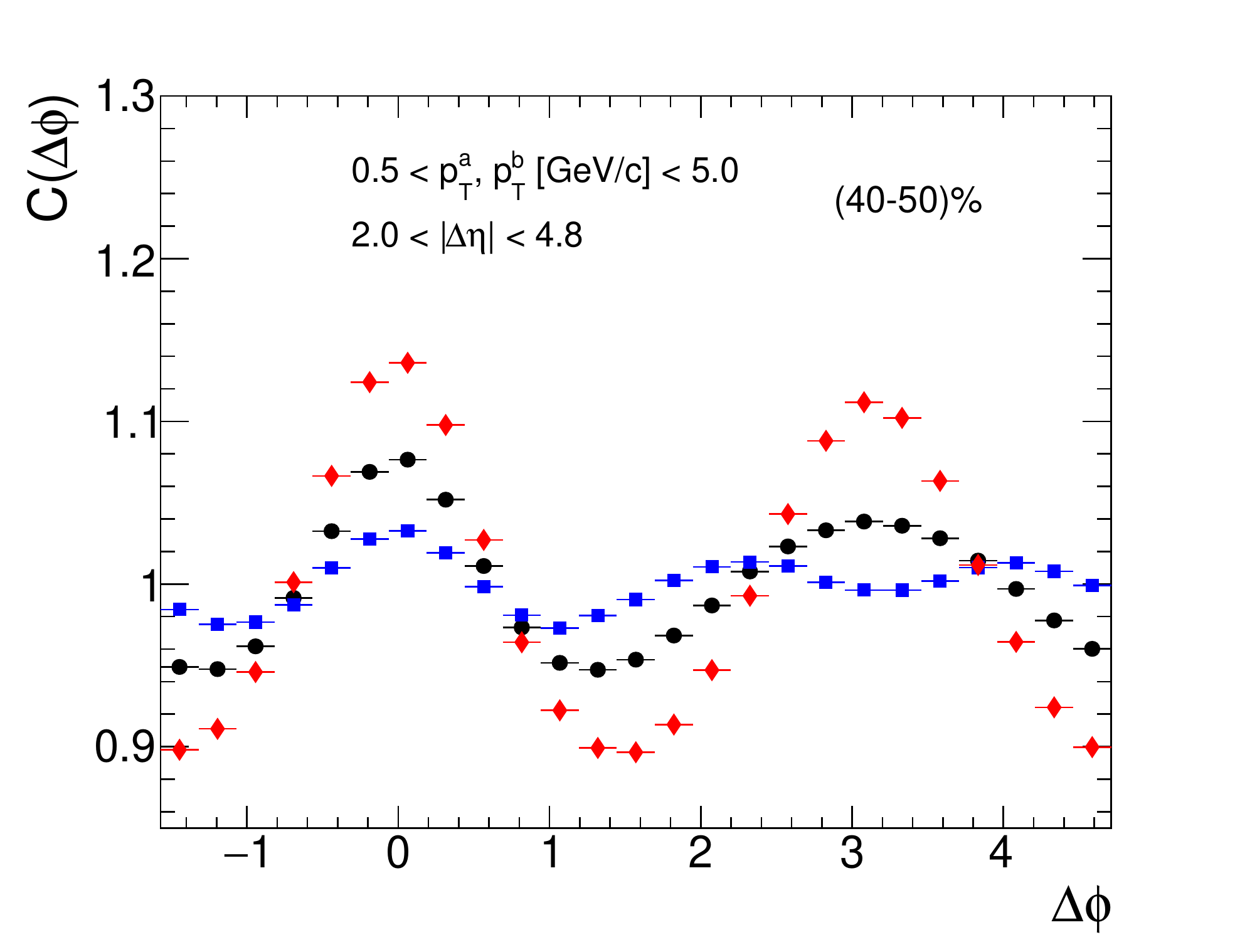}
\caption[width=18cm]{(Color Online) One dimensional two-particle azimuthal correlation function for low-$S_{0}$, high-$S_{0}$ and $S_{0}$ integrated events in Pb--Pb collisions at $\rm{\sqrt{s_{NN}} =  5.02}$  TeV for (0-10)\% (top) and (40-50)\% (bottom) centrality classes \cite{Mallick:2020ium}. }
\label{fig:cdeltaphi}
\end{figure}
\begin{equation}
    C(\Delta\phi)=\frac{dN_{\rm pairs}}{d\Delta\phi}=A\times\frac{S(\Delta\phi)}{B(\Delta\phi)}=A\times\frac{\int S(\Delta\eta,\Delta\phi) d\Delta\eta}{\int B(\Delta\eta,\Delta\phi) d\Delta\eta},
\label{eqn-cdelratio}
\end{equation}
where $A$ is the normalisation constant given as $\rm N_{pairs}^{mixed}/N_{pairs}^{same}$. Here $\rm N_{pairs}^{mixed}$ and $\rm N_{pairs}^{same}$ are mixed event pairs and same event pairs, respectively in the chosen $\Delta\eta$ region. The normalisation scaling in Eq.~\ref{eqn-cdelratio} ensures the same number of pairs in mixed-event (B) and same-event (S). The distribution of pairs or 1D correlation function in $\Delta\phi$ can be Fourier expanded as follows:
\begin{equation}
     C(\Delta\phi)=\frac{dN_{\rm pairs}}{d\Delta\phi} \propto [1+2\sum_{n=1}^{\infty}v_{n,n}(p_{T}^{a},p_{T}^{b})\cos{n\Delta\phi}]
     \label{eqn-cdelfourier}
\end{equation}
where $v_{n,n}$ are the two-particle flow coefficients being symmetric with respect to $p_{T}^{a}$ and $p_{T}^{b}$. $v_{n,n}$ can be calculated by discrete Fourier transformation as:
\begin{equation}
   v_{n,n}(p_{T}^{a},p_{T}^{b})=\langle\cos{n\Delta\phi}\rangle= \frac{\sum_{m=1}^{N}\cos{(n\Delta\phi_{m})}C(\Delta\phi_{m})}{\sum_{m=1}^{N}C(\Delta\phi_{m})}
\end{equation}
where N (=200) is the number of bins in the range $-\pi/2<\Delta\phi<3\pi/2$ of $\Delta\phi$ distribution.

Flow coefficients defined in Eq.~\ref{eqn-invfourierexp} contributes to Eq.~\ref{eqn-cdelfourier} as:

\begin{equation}
    C(\Delta\phi) \propto [1+2\sum_{n=1}^{\infty}v_{n}(p_{T}^{a})v_{n}(p_{T}^{b})\cos{n\Delta\phi}]
    \label{eqn-cdelvn}
\end{equation}
 In Eq.~\ref{eqn-cdelvn}, the event plane angle drops out during the convolution leaving only the dependence on an azimuthal angle. If one assumes that the collective flow is driven by azimuthal anisotropy, then the two-particle harmonic coefficient ($v_{n,n}$) should be a product of two single-particle harmonic coefficients. Therefore,
 \begin{equation}
     v_{n,n}(p_{T}^{a},p_{T}^{b})=v_{n}(p_{T}^{a})v_{n}(p_{T}^{b})
     \label{eqn-vntovnn}
 \end{equation}
 Another way around, we can calculate $v_{n}$ from $v_{n,n}$ by using the following expression
\begin{equation}
    v_{n}(p_{T}^{a})=\frac{v_{n,n}(p_{T}^{a},p_{T}^{b})}{\sqrt{v_{n,n}(p_{T}^{b},p_{T}^{b})}}
    \label{eqn-vnntovn}
\end{equation}

A crucial point to note here that, negative value of $v_{n,n}(p_{T}^{b},p_{T}^{b})$ in Eq.~\ref{eqn-vnntovn} makes $v_{n}(p_{T}^{a})$ imaginary which is not physical.
The correlation function and the anisotropic flow coefficients such as elliptic and triangular flow are calculated in the 
pseudorapidity range $|\eta| < 2.5$, with a relative pseudorapidity gap of $2 < |\Delta\eta| < 4.8$ in the transverse momentum range of the particle pairs from 0.5 to 5 GeV/c \cite{Mallick:2020ium}. Figure \ref{fig:cdeltaphi} shows the transverse spherocity dependence on one dimensional two-particle azimuthal correlation function plotted with respect to $\Delta\phi$ for (0-10)$\%$ and (40-50)$\%$ centralities in Pb--Pb collisions at $\rm{\sqrt{s_{NN}} =  5.02}$ TeV. The larger correlation amplitude for mid-central collisions compared to the central collisions indicates more azimuthal anisotropy in mid-central than the central collisions. There is a strong dependence of the correlation function on the spherocity for any particular centrality. One can infer from Fig.~\ref{fig:cdeltaphi} that low-$S_0$ events have more azimuthal anisotropy than high-$S_0$ events. Two peaks in the away side in $\Delta\phi$ for high-$S_0$ events manifest due to higher contribution from $v_3$.

\begin{figure}[ht!]
\begin{center}
\includegraphics[scale=0.42]{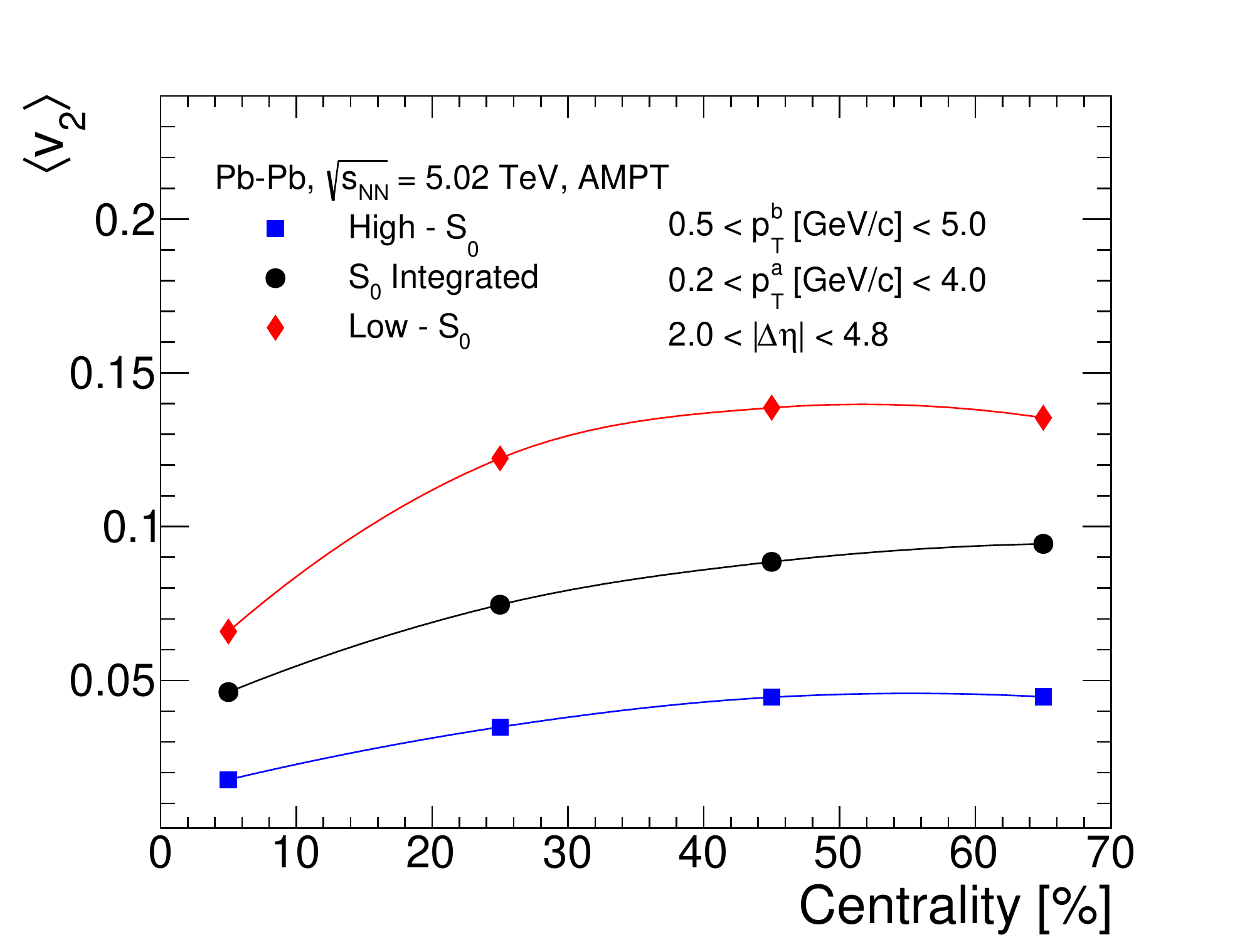}
\caption[width=18cm]{(Color Online) Mean elliptic flow ($\langle v_2\rangle$) vs centrality for different spherocity classes in Pb--Pb collisions at $\rm \sqrt{s_{NN}}$ = 5.02 TeV.}
\label{fig:v22andv2}
\end{center}
\end{figure}

Figure \ref{fig:v22andv2} represents centrality dependence of elliptic flow ($\langle v_2\rangle$) for different spherocity classes in Pb--Pb collisions at $\rm{\sqrt{s_{NN}} =  5.02}$ TeV. In high-energy heavy-ion collisions, due to the presence of a pressure gradient formed in the medium, anisotropy in the geometry of the initial nuclear overlap region is transformed into final state azimuthal anisotropy. Since central collisions are almost isotropic in geometry, corresponding $v_2$ is also less. However, if one moves towards mid-central collisions, the elliptic flow coefficient increases since the eccentricity is higher. However, in peripheral collisions, although eccentricity is remarkably high, due to the lack of the number of participants a smaller size and shorter lifetime of the fireball, spatial anisotropy can not be completely transformed into $v_2$. Figure~\ref{fig:v22andv2} also shows that the value of elliptic flow significantly depends on event selection-based transverse spherocity. Low-$S_0$ events have higher $\langle v_2\rangle$ while we observe almost negligible $\langle v_2\rangle$ in high-$S_0$ events.

\begin{figure}[ht!]
\begin{center}
\includegraphics[scale=0.42]{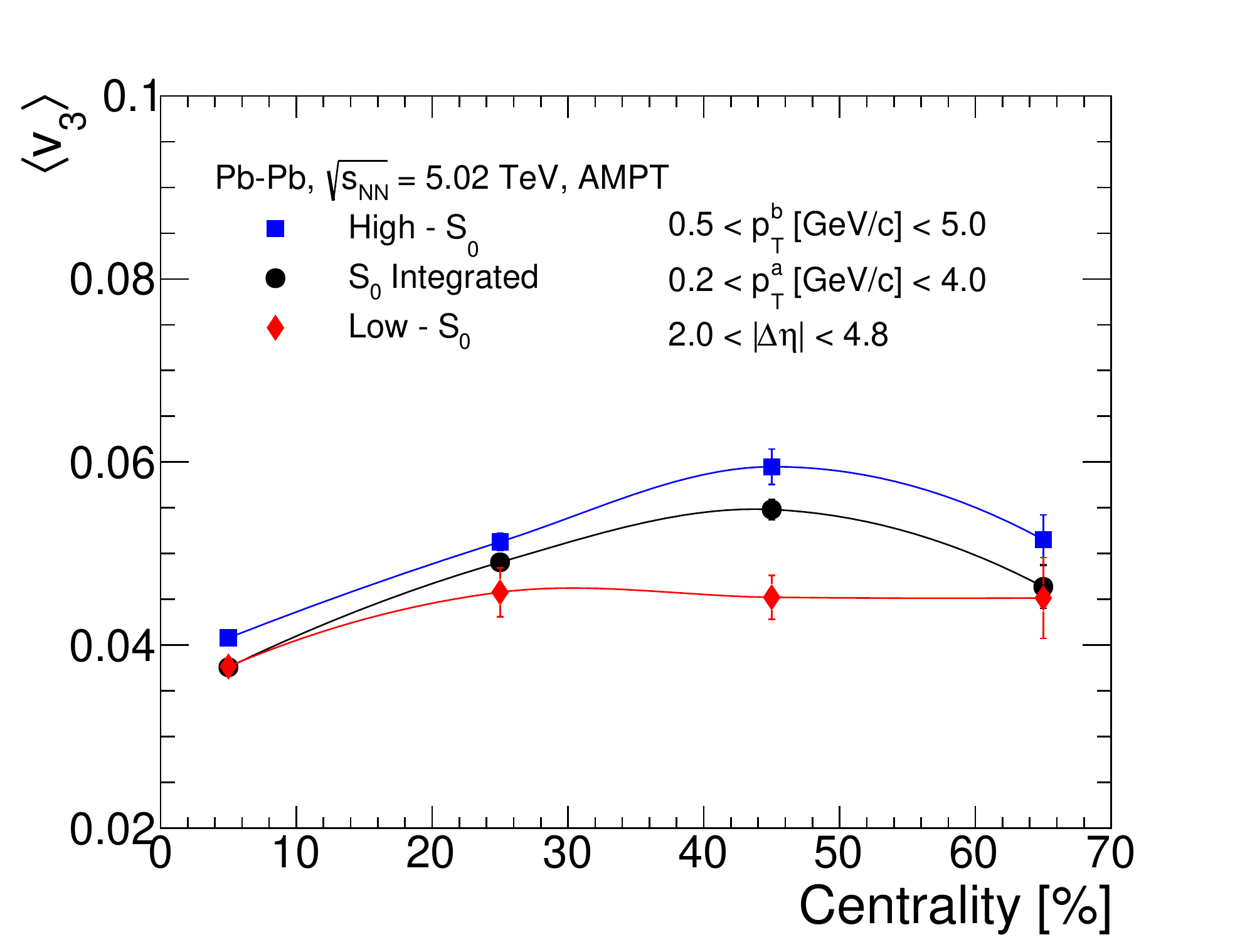}
\caption[width=18cm]{(Color Online) Mean triangular flow ($\langle v_{3}\rangle$) as a function of centrality for different spherocity classes in Pb--Pb collisions at $\rm \sqrt{s_{NN}}$ = 5.02 TeV.}
\label{fig:v33andv3}
\end{center}
\end{figure}

Figure~\ref{fig:v33andv3} shows triangular flow ($\langle v_3\rangle$) as a function of centrality for different spherocity classes. In mid-central collisions, one observes slightly higher $v_3$ than the central collisions for a specific spherocity class. The dependence of $v_3$ on centrality is weaker than that of $v_2$, which can also be observed from Figs. \ref{fig:v22andv2} and \ref{fig:v33andv3}. This weak dependence of $v_3$ on centrality is also observed in experiments \cite{ATLAS:2015qwl,ALICE:2011ab}. 
In addition, one can observe appreciable dependence of triangular flow on transverse spherocity. In Fig.~\ref{fig:v22andv2}, $v_2$ contribution is maximum in case of low-$S_0$ events while in Fig.~\ref{fig:v33andv3} the trend is reversed and high-$S_0$ events have dominating $v_3$. This anti-correlation between $v_2-v_3$ for $q_2$-selection as pointed out in Ref. \cite{ATLAS:2015qwl} can also be deduced from this study with transverse spherocity. However the anti-correlation between $\varepsilon_2-\varepsilon_3$ is not observed in Fig.~\ref{fig:ratioe2e3}. One may infer that the source of this anti-correlation between $v_2-v_3$ with spherocity selection may not be propagated from the initial geometry but may have effects from the medium during its evolution due to the fact that the fluidity of the medium affects differently to different flow coefficients. This is an indication that the fluidity of the medium is spherocity-dependent. In Ref.~\cite{Chaudhuri:2011pa}, authors have shown the dependence of specific shear viscosity ($\eta/s$) on the correlation between triangularity and triangular flow, where the correlation decreases with an increase in $\eta/s$ and the correlation between $v_3$ and $\varepsilon_3$ is not strong in an ideal or minimally viscous fluid. Only (65-70)\% of the triangular flow is related to initial triangularity, conveying that a substantial part of the triangular flow is unrelated to the initial triangularity. Although the exact reason for this effect is not yet completely understood; however, effects from higher moments or products of moments (via $v_1-v_2$ coupling) may lead to such observations \cite{Chaudhuri:2011pa}. This may be the reason for the observed dependence of triangular flow on transverse spherocity, while triangularity of the overlap region does not have any spherocity dependence. This can be investigated in a future study as a function of spherocity and it is not covered in the manuscript.

\begin{figure}[ht!]
\begin{center}
\includegraphics[scale=0.4]{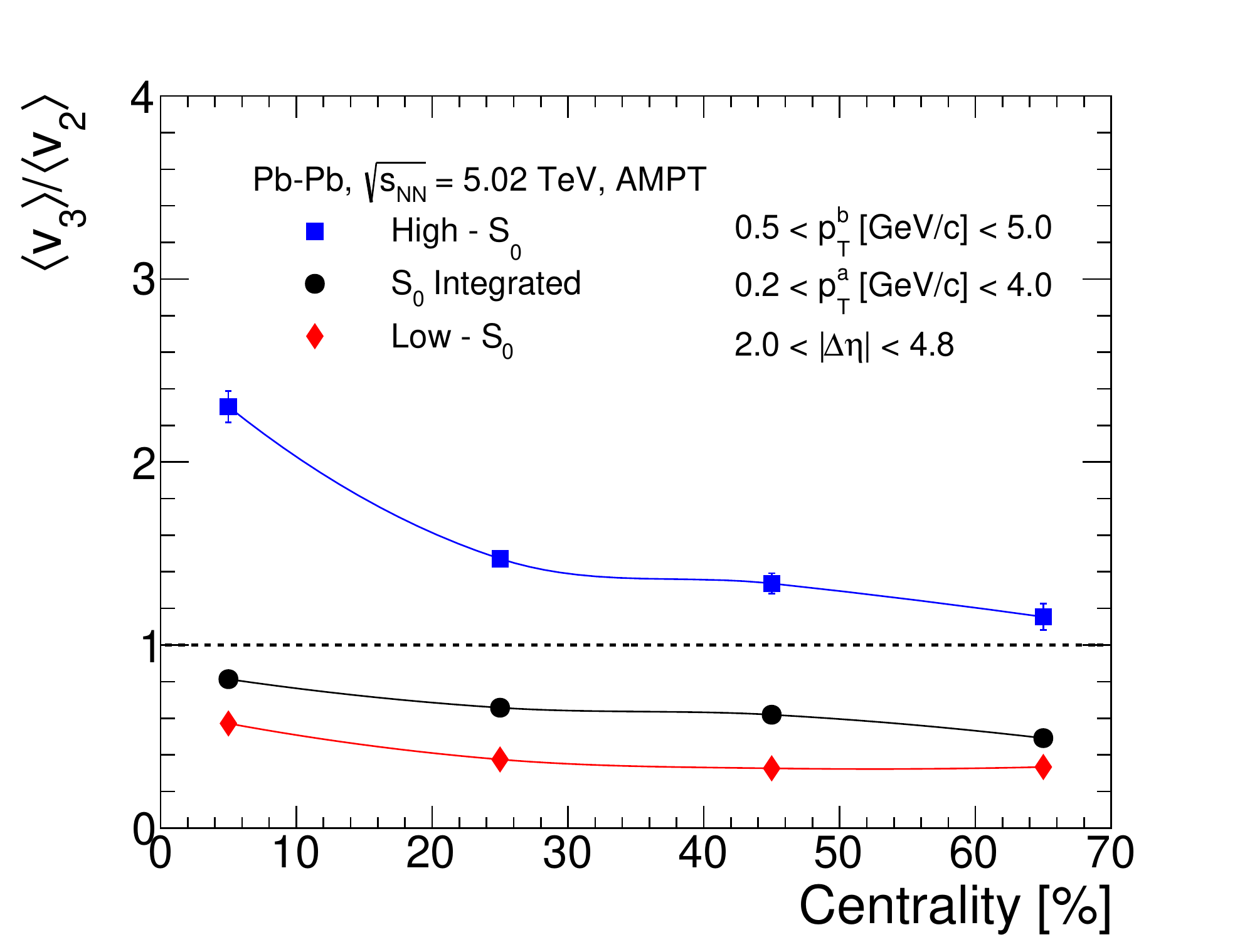}
\caption[width=18cm]{(Color Online) $\langle v_3\rangle/\langle v_2\rangle$ vs centrality for different spherocity events in Pb--Pb collisions at $\sqrt{s_{NN}}$ = 5.02 TeV.}
\label{fig:v3v2ratio}
\end{center}
\end{figure}

Figure~\ref{fig:v3v2ratio} represents the ratio $\langle v_3\rangle/\langle v_2\rangle$ versus centrality for different spherocity events. One notices that $\langle v_3\rangle/\langle v_2\rangle$ is decreasing when going from central to peripheral collisions and shows considerable spherocity dependence. This may be because of three reasons; due to the reminiscent centrality dependence of $\langle \varepsilon_3\rangle/\langle \varepsilon_2\rangle$ and/or due to viscous effects \cite{ALICE:2011ab}, or have a contribution from both the effects. One observes $\langle \varepsilon_3\rangle/\langle \varepsilon_2\rangle$ vs centrality trend is followed by $\langle v_3\rangle/\langle v_2\rangle$ starting from most central until semi central collisions. Here, with small changes in centrality, the viscous effect might cause a little change in $\langle v_3\rangle/\langle v_2\rangle$, however, for the peripheral collisions, the viscous effects play a major role and $\langle \varepsilon_3\rangle/\langle \varepsilon_2\rangle$ trend is not followed by $\langle v_3\rangle/\langle v_2\rangle$ for all spherocity classes. One can notice that $\langle v_3\rangle/\langle v_2\rangle$ is highest and always greater than one for high-$S_0$ events showing the dominance of $v_3$ over $v_2$ for all centrality classes and the low-$S_0$ curve shows the dominance of $v_2$ over $v_3$. This effect is expected to have propagated from initial geometry, which is evident from $\langle \varepsilon_3\rangle/\langle \varepsilon_2\rangle$ plot shown in Fig.~\ref{fig:ratioe2e3}.

\begin{figure}[ht!]
\begin{center}
\includegraphics[scale=0.4]{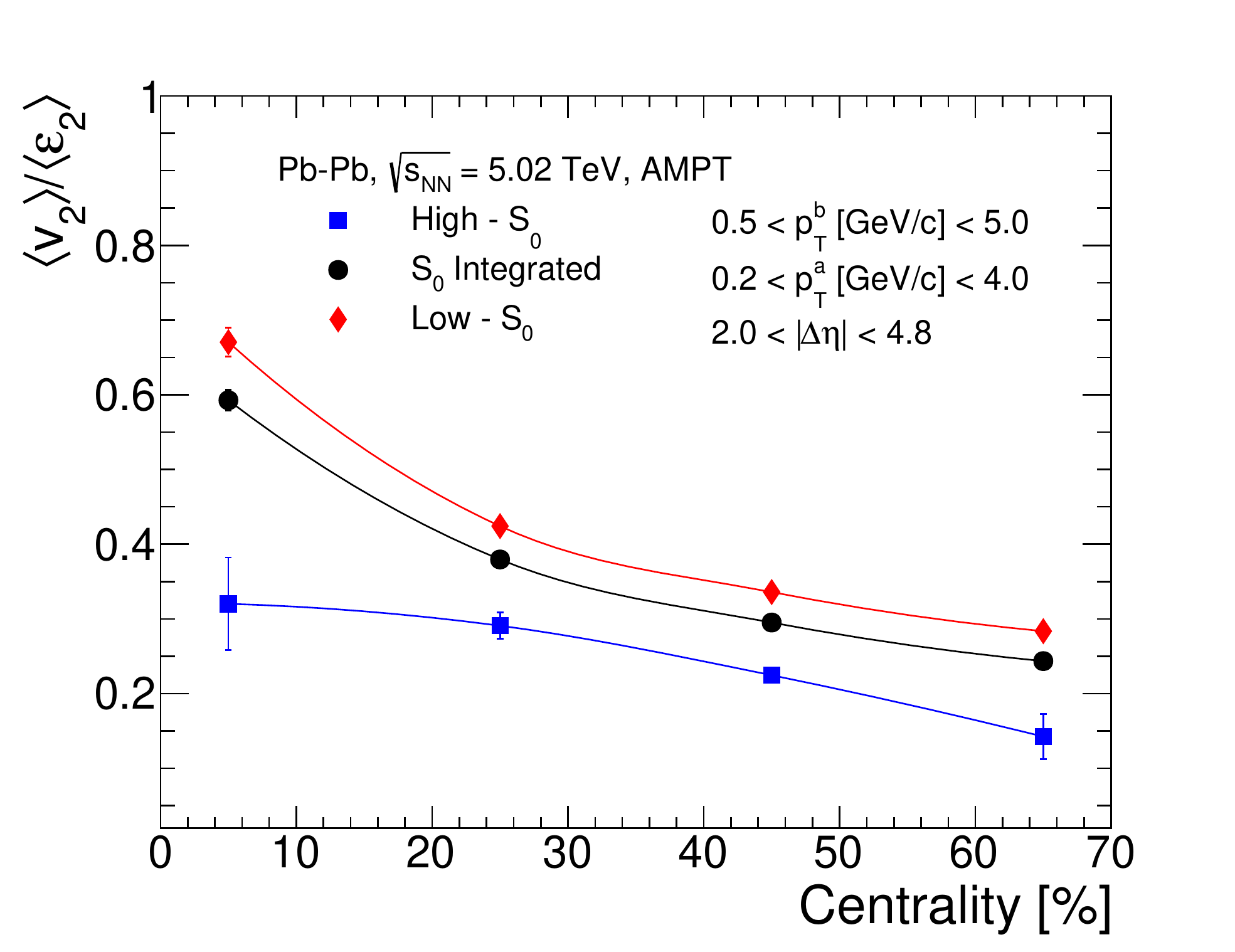}
\includegraphics[scale=0.4]{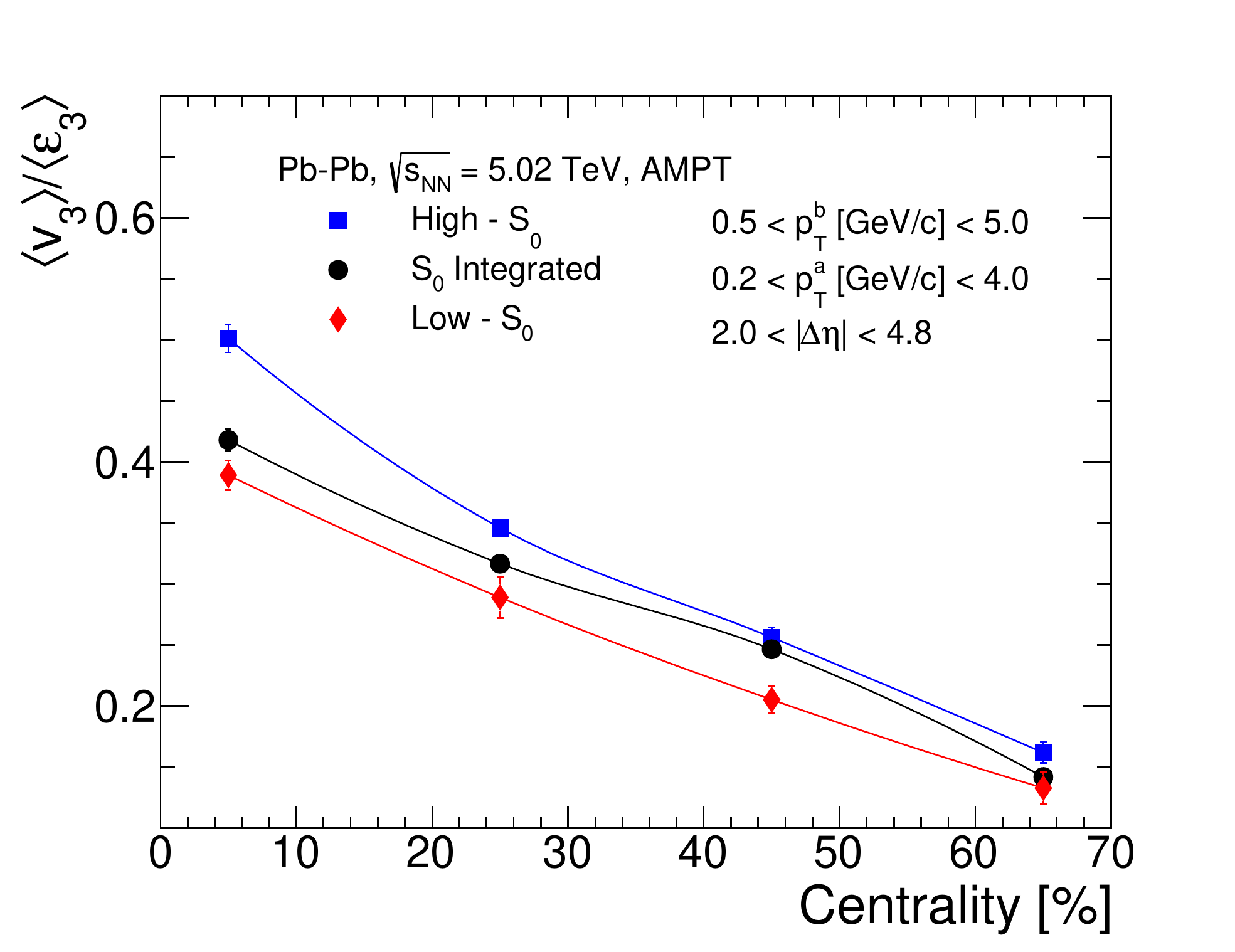}
\caption[width=18cm]{(Color Online) $\langle v_2 \rangle/\langle \epsilon_2 \rangle$ (top) and $\langle v_3 \rangle/\langle \epsilon_3 \rangle$ (bottom) vs centrality for different spherocity classes in Pb--Pb collisions at $\rm\sqrt{s_{NN}}$ = 5.02 TeV using AMPT.}
\label{fig:vnenvscent}
\end{center}
\end{figure}

Figure~\ref{fig:vnenvscent} represents the elliptic and triangular flow normalised with eccentricity and triangularity, respectively, as a function of centrality in Pb--Pb collisions at $\rm\sqrt{s_{NN}}$ = 5.02 TeV using AMPT. The figure qualitatively tells about the response of the medium formed, and its evolution to different centrality and spherocity selections. In a perfect fluid, $v_n\propto\epsilon_n$ and thus $v_n/\epsilon_n$ is expected to be a constant value. However, in Fig.~\ref{fig:vnenvscent}, both $\langle v_2 \rangle/\langle \epsilon_2 \rangle$ and $\langle v_3 \rangle/\langle \epsilon_3 \rangle$ are found to be varying with spherocity and centrality. As we move from central to peripheral collisions, $\langle v_n \rangle/\langle \epsilon_n \rangle$ decreases. This trend with respect to different centrality selection is also observed in experimental results \cite{ALICE:2011ab}.

\begin{figure}[ht!]
\begin{center}
\includegraphics[scale=0.4]{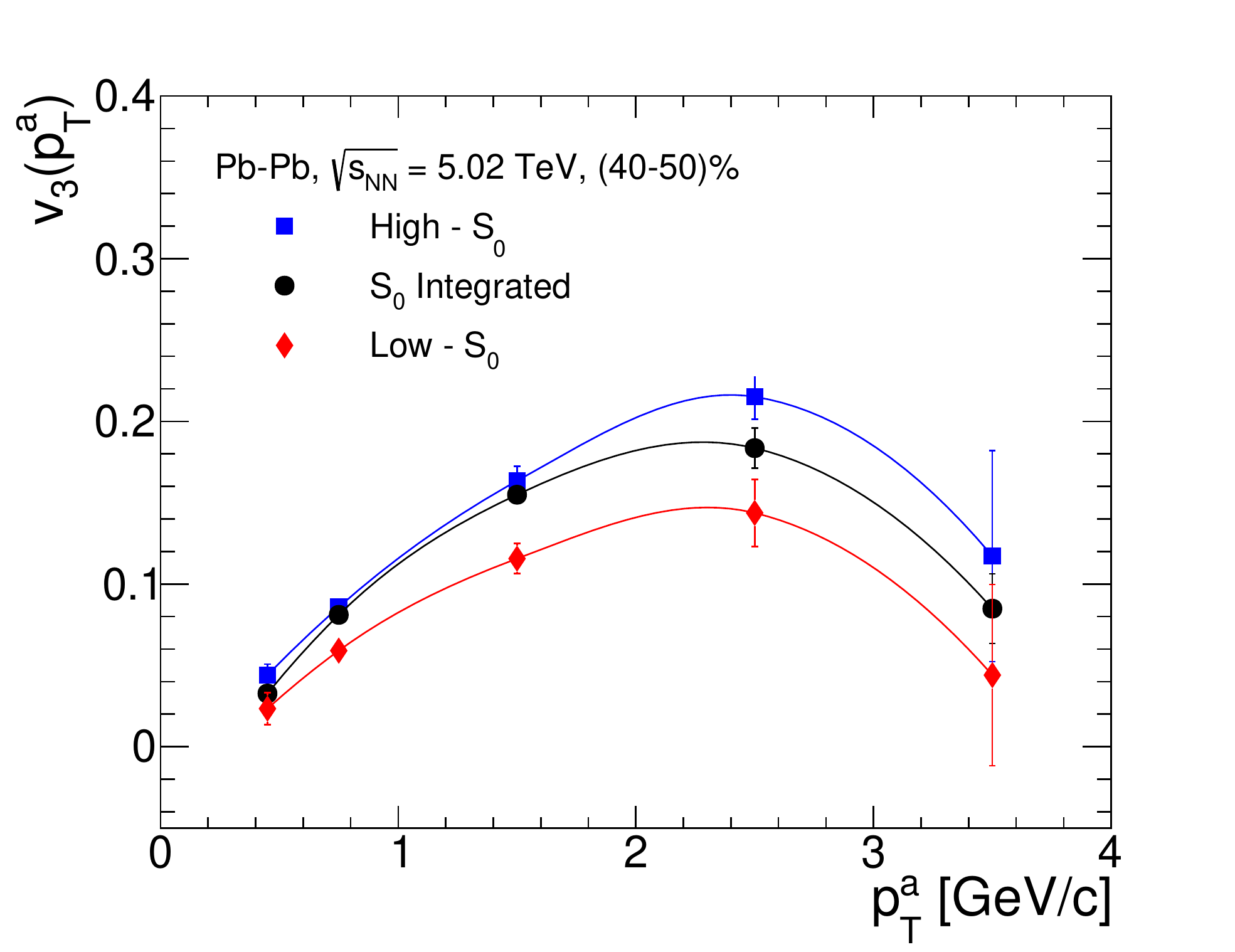}
\caption[width=18cm]{(Color Online) Single particle triangular flow ($v_{3}(p_{\rm T}^{\rm a})$)  for high-$S_{0}$, $S_{0}$ integrated and low-$S_{0}$ events for Pb--Pb collisions at $\sqrt{s_{NN}} = $ 5.02 TeV for (40-50)\% centrality class.}
\label{fig:v33v3}
\end{center}
\end{figure}

Figure~\ref{fig:v33v3} shows single particle triangular flow ($v_{3}(p_{\rm T}^{\rm a})$) as a function of $p_{\rm T}$ for different centrality classes for (40-50)\% centrality in Pb--Pb collisions at $\sqrt{s_{NN}} = $ 5.02 TeV. $v_{3}(p_{\rm T}^{\rm a})$ shows an increase with $p_{\rm T}$ from low to intermediate $p_{\rm T}$ region, attains a maximum, and starts decreasing towards higher $p_{\rm T}$. Similar trend is observed in Ref. \cite{Mallick:2020ium} for $v_{2, 2}(p_{\rm T}^{\rm a}, p_T^b)$ and $v_{2}(p_{\rm T}^{\rm a})$. One observes a remarkable dependence of transverse spherocity on the triangular flow when compared to elliptic flow. High-$S_0$ events have the highest $v_{3}(p_{\rm T}^{\rm a})$ value, whereas low-$S_0$ events have the lease $v_{3}(p_{\rm T}^{\rm a})$ value.

\begin{figure*}[ht!]
\begin{center}
\includegraphics[scale=0.29]{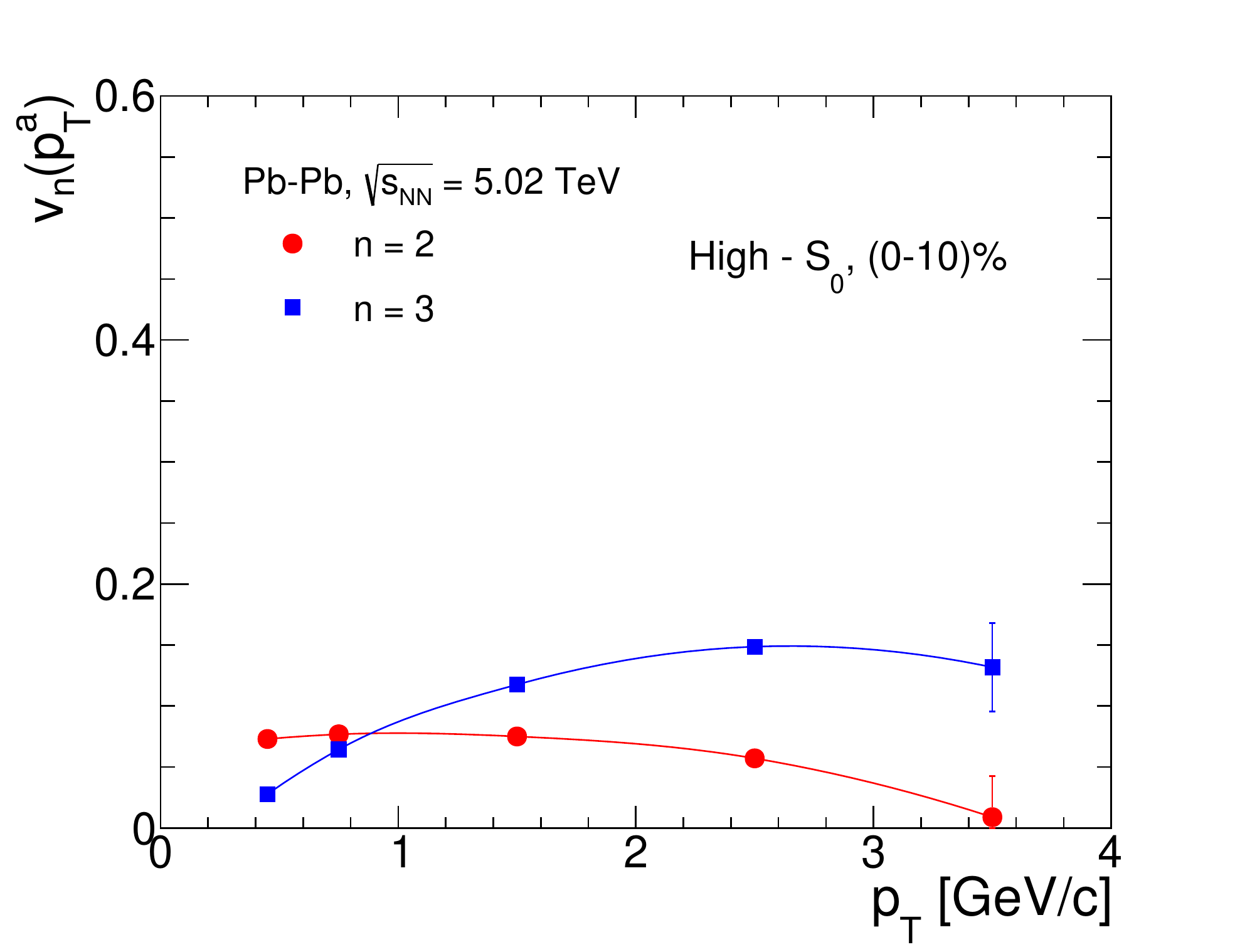}
\includegraphics[scale=0.29]{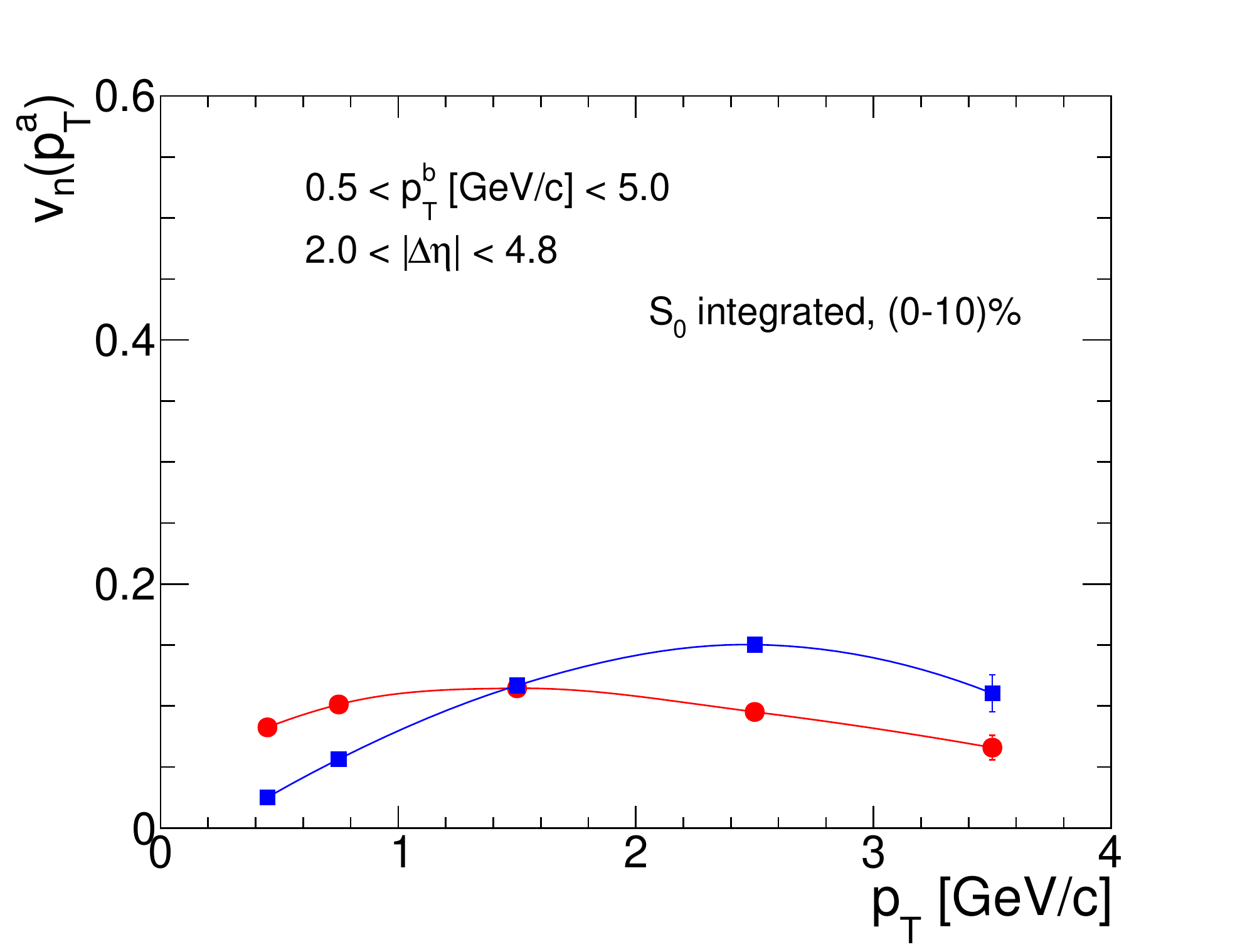}
\includegraphics[scale=0.29]{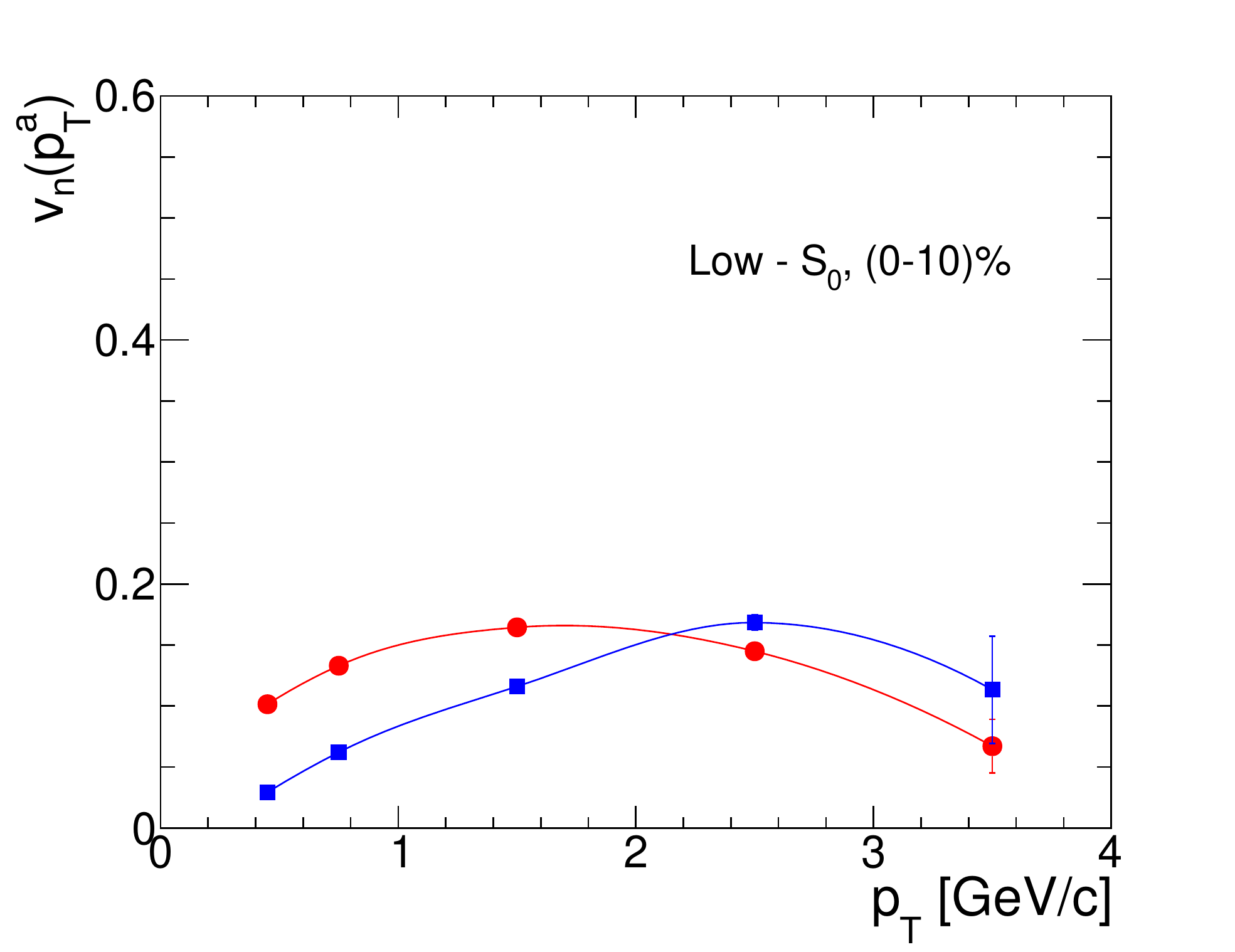}
\includegraphics[scale=0.29]{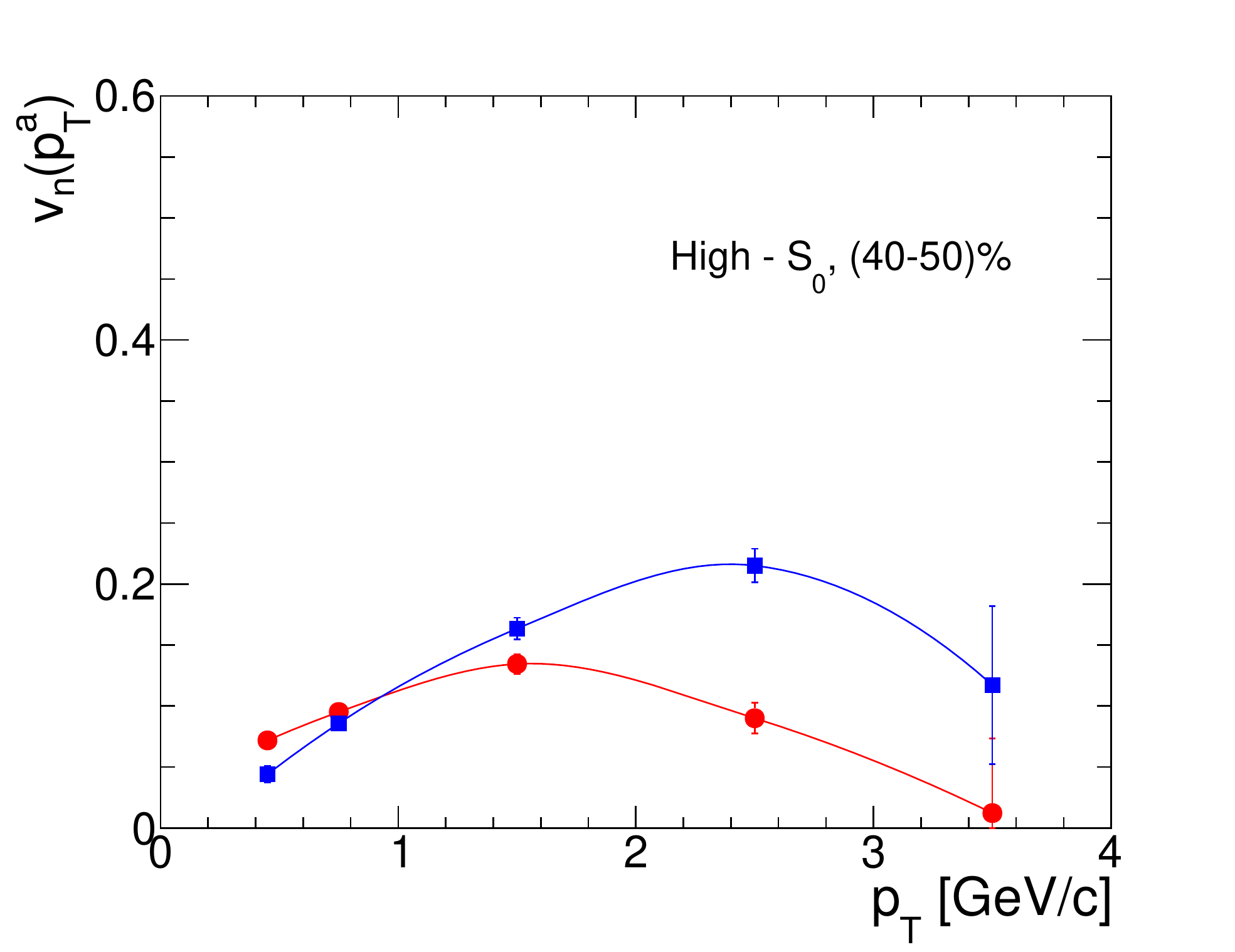}
\includegraphics[scale=0.29]{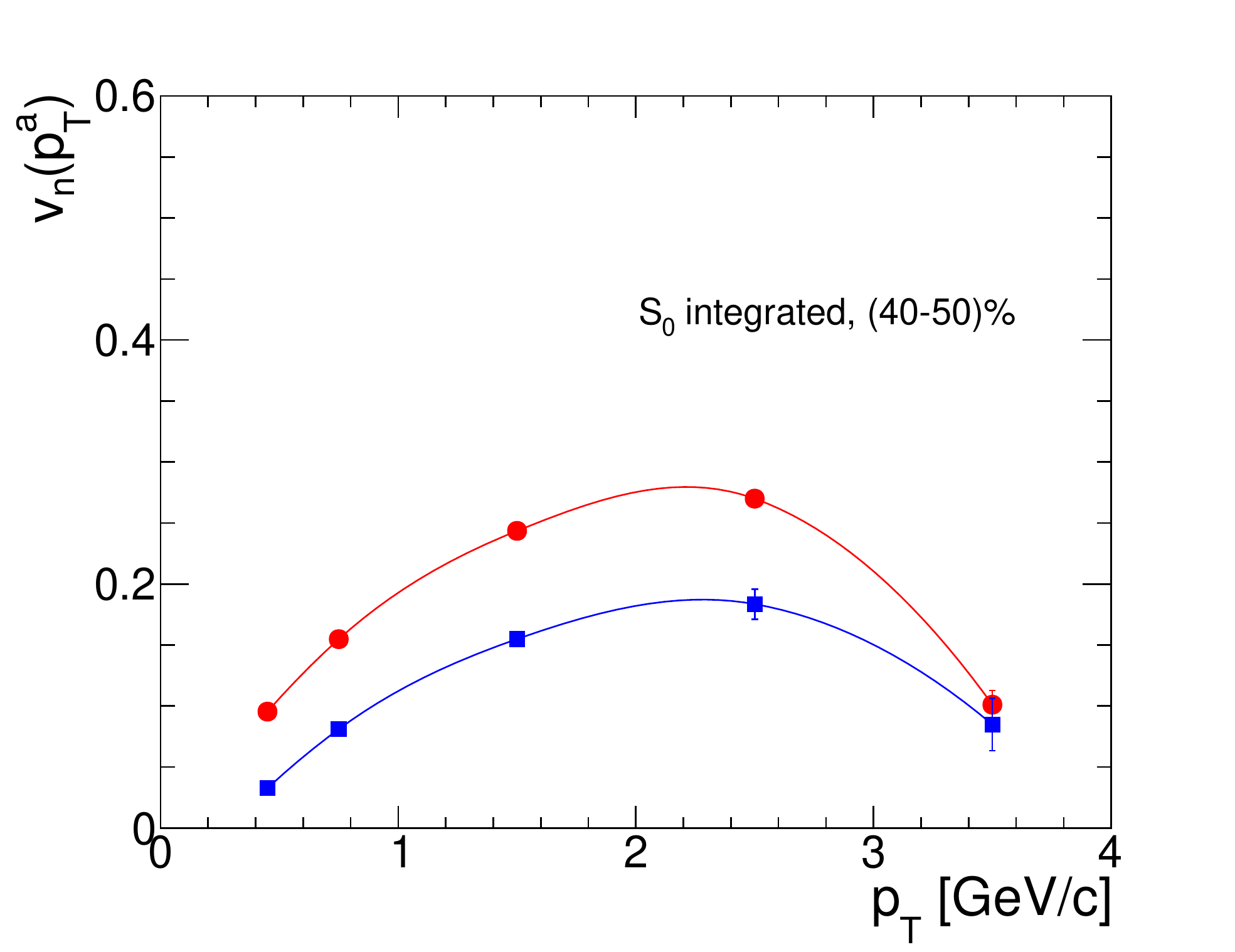}
\includegraphics[scale=0.29]{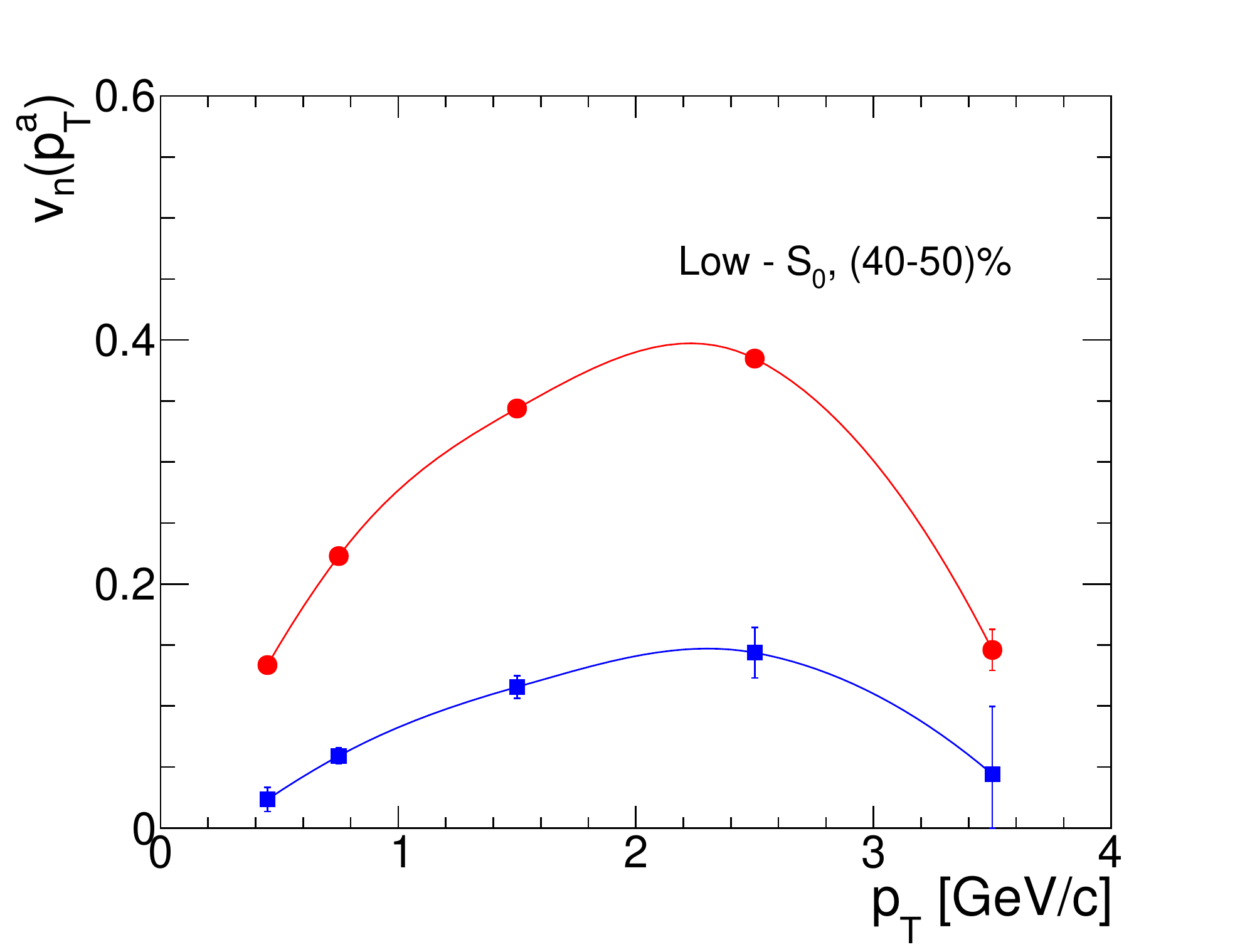}
\caption[width=18cm]{(Color Online) Single particle elliptic (red) and triangular (blue) flow for high-$S_{0}$ (left column), $S_{0}$ integrated (middle column) and low-$S_{0}$ (right column) events for Pb--Pb collisions at $\sqrt{s_{NN}} = $ 5.02 TeV for (0-10)\% (top) and (40-50)\% (bottom) centrality classes using AMPT.}
\label{fig:crossingpt}
\end{center}
\end{figure*}

\begin{figure}[ht!]
\begin{center}
\includegraphics[scale=0.4]{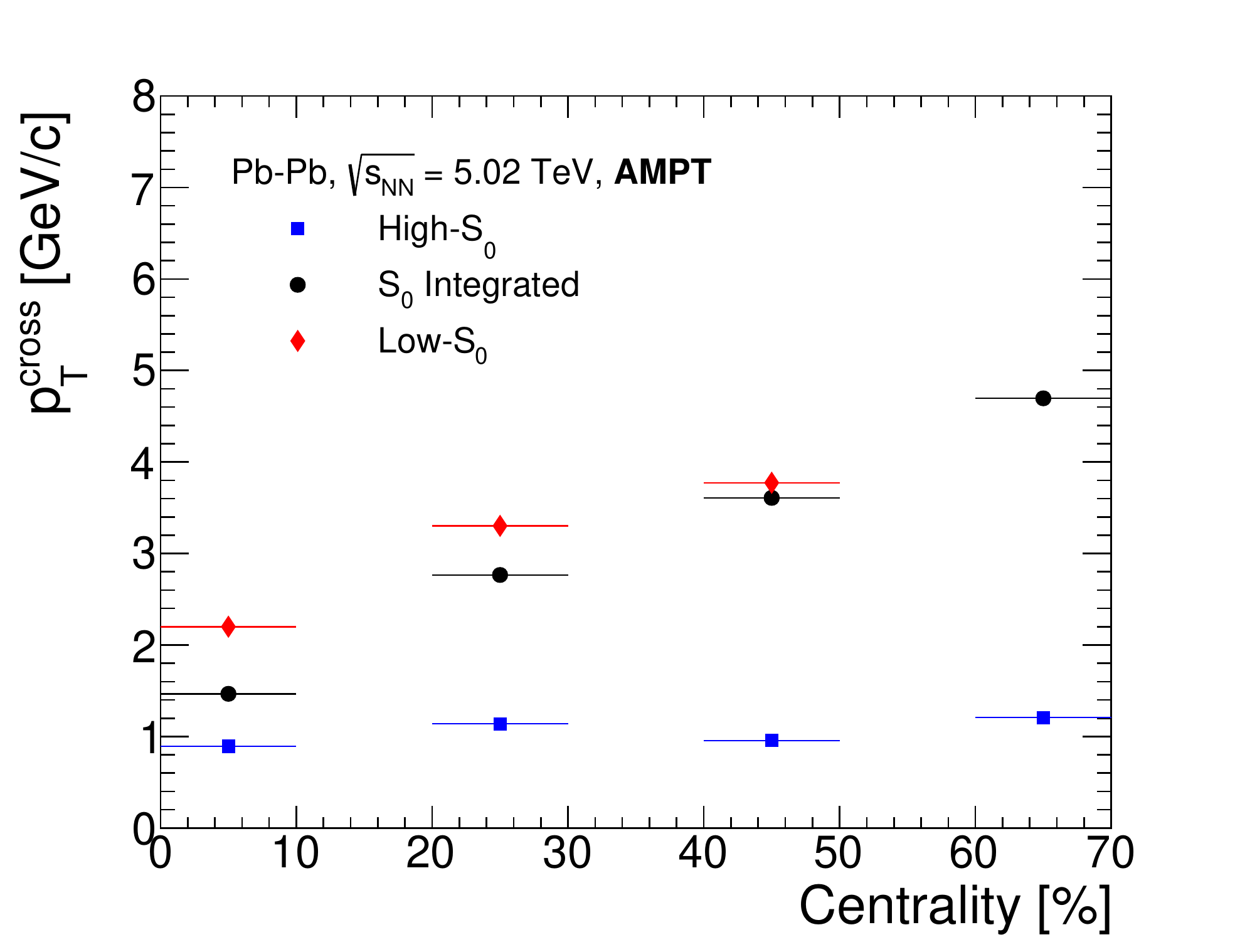}
\caption[width=18cm]{(Color Online) Transverse momentum value corresponding to crossing between $v_2$ and $v_3$ ($\rm p_T^{cross}$) as a function of centrality for different spherocity cuts for Pb--Pb collisions at $\sqrt{s_{NN}}$ = 5.02 TeV using AMPT.}
\label{fig:crossingvscent}
\end{center}
\end{figure}

Figure~\ref{fig:crossingpt} shows the comparison between elliptic and triangular flow as a function of $p_{\rm T}$ for (0-10)\% and (40-50)\% centrality for different spherocity classes. For the mid-central collisions, $v_2$ is higher than $v_3$ except for the high-$S_0$ case. However, in the most central case we observe the domination of $v_3$ over $v_2$ after a certain $p_{\rm T}$ value. The crossing point in $p_{\rm T}$ ($p_{\rm T}^{\rm cross}$), where the values of $v_2$ and $v_3$ become equal, appear to change for different centrality classes. This behaviour is also observed and is in agreement with experimental data \cite{ALICE:2011ab,ALICE:2016cti,ALICE:2018yph}, where $p_{\rm T}^{\rm cross}$ is found to be increasing with particles' mass and centrality. As pointed out in Refs. \cite{ALICE:2016cti,ALICE:2018yph}, the crossing point between the flow coefficients is attributed to the interplay of these flow coefficients with the radial flow \cite{ALICE:2018yph}. This is because, for central collisions, the contribution of fluctuations in the initial nuclear distribution is more than the influence of the overlap region on the development of the anisotropic flow. However, in the peripheral collisions, the collision geometry contributes higher than initial density fluctuations \cite{ALICE:2016cti}. In this work, we go a step ahead and try to see the dependence of $p_{\rm T}^{\rm cross}$ with event topology which is shown explicitly in Fig.~\ref{fig:crossingvscent}. $p_{\rm T}^{\rm cross}$ for different centrality  and spherocity classes is extracted by fitting a polynomial function to the plots in Fig.~\ref{fig:crossingpt}. $p_{\rm T}^{\rm cross}$ is found to be almost flat for high-$S_0$ events but it is observed to be increasing with centrality for $S_0$ integrated and low-$S_0$ cases. The expected low $p_{\rm T}^{\rm cross}$ for high-$S_0$ events can be accounted for due to a higher contribution from the initial density fluctuations than the influence of initial collision geometry on the anisotropic flow as shown in Fig.~\ref{fig:ratioe2e3}.

\section{Summary}
\label{section4}

In this paper, we have explored the eccentricity, triangularity, elliptic flow, and triangular flow along with their correlations in Pb--Pb collisions at $\rm\sqrt{s_{NN}}$ = 5.02 TeV in the framework of a multi-phase transport model using event shape engineering tools such as the transverse
spherocity. The important findings are summarised below:

\begin{itemize}
    
    \item After its successful implementation in small collision systems, in this work, we found a significant correlation of transverse spherocity with the more widely used event shape classifier, the reduced flow vector. This highlights the advantage of using spherocity as a unique event shape classifier across all collision systems at the LHC. 

    \item Since the eccentricity is found to be varying with spherocity selection, elliptic flow is found to be strongly (anti-)correlated with spherocity selection as well. As opposed to the initial triangularity, triangular flow shows a significant dependence on transverse spherocity.
    
    \item Through the studies using the Pearson coefficient, we found eccentricity and triangularity show a relatively higher degree of correlation for high-$S_0$ events for all the centrality classes, as compared to the low-$S_0$ and spherocity-integrated events.

    \item We report an increase in $\langle v_3\rangle/\langle v_2\rangle$ towards central collisions, and the ratio is always greater than one for high-$S_0$ events. This is expected to be propagated from the initial geometry of the participant nucleons and may have contributions from the medium formed.
    \item We report a crossing point between $v_2$ and $v_3$ at a certain transverse momentum value ($\rm p_T^{cross}$) which is found to be varying with centrality and transverse spherocity. $p_{\rm T}^{\rm cross}$ is found to be decreasing when going towards either central or high-$S_0$ events.
\end{itemize}

The observables related to the geometry of the nuclear overlap region and the anisotropic flow coefficients are found to be correlated among themselves as well as with transverse spherocity. The anisotropic flow coefficients are expected to have a contribution from the medium formed in heavy-ion collisions. The present event topological study using the AMPT transport model gives us clues of new findings, which are yet to be verified in experimental data. These event shape studies including
the small systems would be more interesting to disentangle initial versus final state effects in the discussed observables.
This study paves a new way of understanding the medium formation through event topology in heavy-ion collisions.
So far, there have been no studies with event-shape observables both in small and large collision systems at the LHC. Using the same event classifier in both large and small systems is very important due to the recent discoveries of QGP-like effects in small systems.  In this paper, we show the feasibility of using the transverse spherocity as an event shape observable in heavy-ion collisions, and along with the previous successful use of it in small systems, one can understand the possible reasons for the QGP-like effects in small systems and the associated particle production dynamics.

\section*{Appendix}
\subsection{Components of AMPT model}
\label{AMPTComp}
\begin{enumerate}
    \item Initialization of collisions: This step in AMPT model is performed using HIJING \cite{Wang:1991hta}, where a differential cross-section of the produced mini-jets in pp collisions are calculated, and produced partons calculated in pp collisions are converted into A-A and p-A collisions by incorporating parameterised shadowing function and nuclear overlap function using inbuilt Glauber Model.
    
    \item Parton transport: Transportation of produced particles is done using Zhang's Parton Cascade (ZPC) model \cite{Zhang:1997ej}. In the String Melting version of AMPT (AMPT-SM), colored strings melt into low-momentum partons.
    
    \item Hadronization: In AMPT-SM, transported partons are hadronised using spatial coalescence mechanism \cite{Lin:2001zk,He:2017tla}. In the default AMPT version, a fragmentation mechanism using Lund fragmentation parameters is used to hadronize the transported partons.
    
    \item Hadron transport: The produced hadrons undergo final evolution in relativistic transport mechanism through meson-meson, meson-baryon, and baryon-baryon interactions \cite{Li:2001xh,Greco:2003mm}.
    
\end{enumerate}

\subsection{Comparison of results from AMPT model and experimental data}
\begin{figure}[ht!]
\includegraphics[scale=0.42]{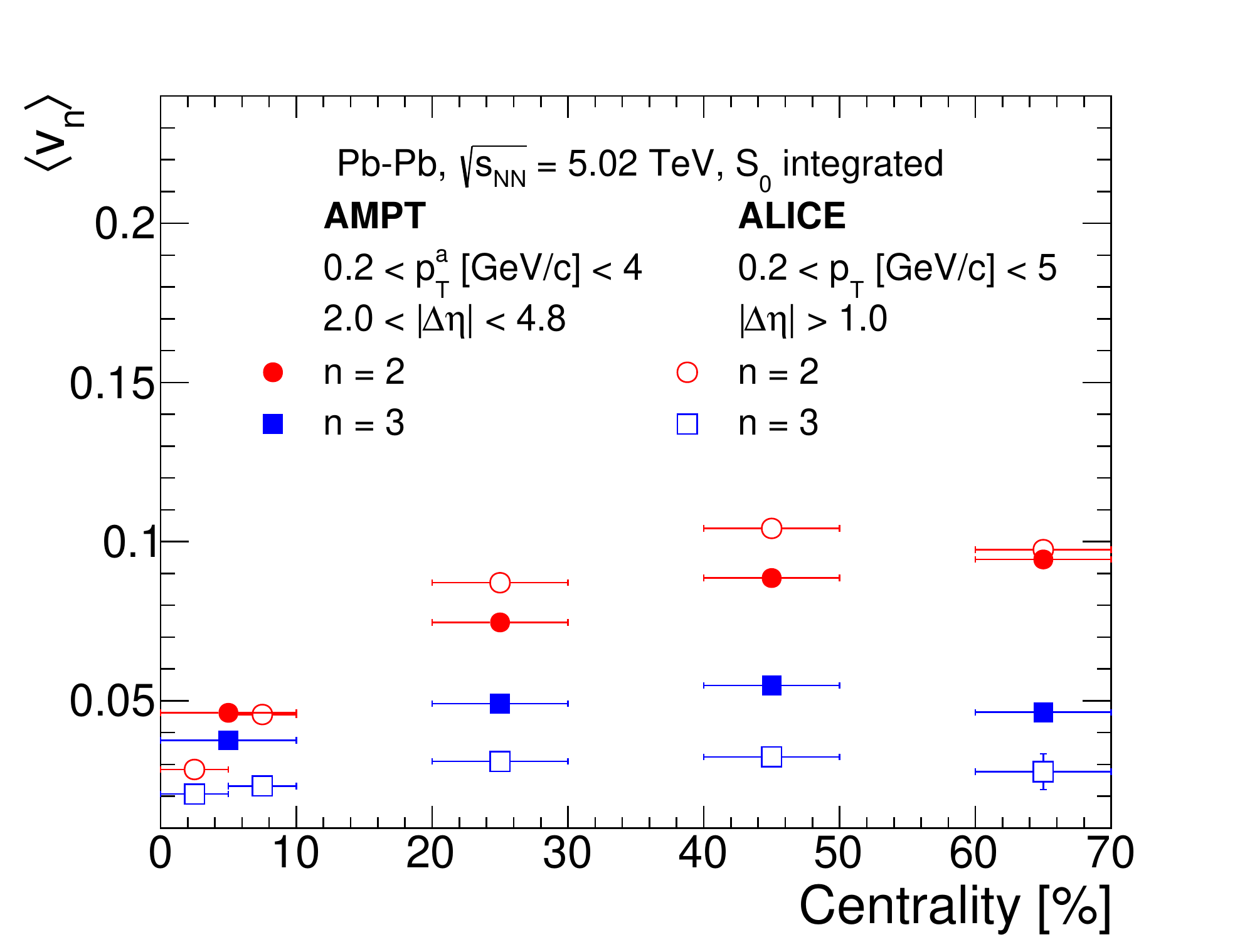}
\includegraphics[scale=0.42]{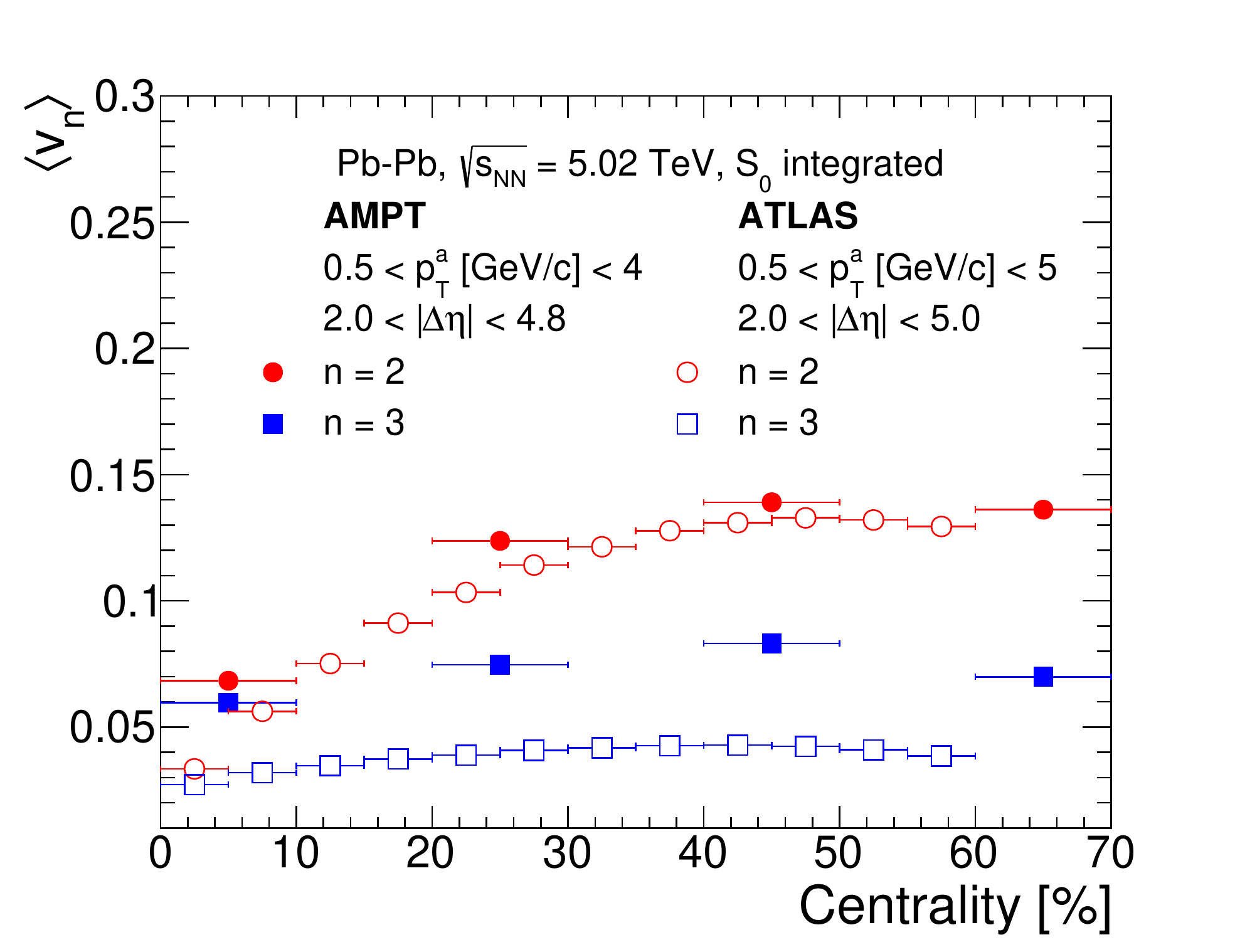}
\caption[width=18cm]{(Color Online) $\langle v_2\rangle$ and $\langle v_3\rangle$ vs centrality for $S_0$ integrated events for Pb--Pb collisions at $\rm\sqrt{s_{NN}}$ = 5.02 TeV using AMPT compared with the similar ALICE \cite{ALICE:2016ccg} (top) and ATLAS \cite{ATLAS:2019dct} (bottom) results.}
\label{fig:vncompALICE}
\end{figure}

Figure~\ref{fig:vncompALICE} shows the comparison of $\langle v_2\rangle$ and $\langle v_3\rangle$ vs centrality for $S_0$-integrated events for Pb--Pb collisions at $\rm\sqrt{s_{NN}}$ = 5.02 TeV using AMPT with ALICE \cite{ALICE:2016ccg} (top) and ATLAS \cite{ATLAS:2019dct} (bottom) results. AMPT is found to be slightly underestimating the elliptic flow from ALICE and ATLAS, however, overestimates the triangular flow. This disagreement between AMPT and experimental data can be fixed using different settings available in AMPT, which is out of the scope of this paper.

\section*{Acknowledgements}
SP acknowledges the financial support from UGC, the Government of India. ST acknowledges the support under the INFN
postdoctoral fellowship. RS acknowledges the financial support under DAE-BRNS Project No. 58/14/29/2019-BRNS of the Government of India.   The authors would like to acknowledge the usage of resources of the LHC grid Tier-3 computing facility at IIT Indore under the mega-science project grant of DAE-DST, Govt. of India  – “Indian participation in the ALICE experiment at CERN” bearing Project No. SR/MF/PS-02/2021-IITI (E-37123).


\end{document}